\documentclass{iacrtrans}

\usepackage{cite}
\usepackage[utf8]{inputenc}
\usepackage{enumitem}
\usepackage{amsmath}
\usepackage{mathtools}
\usepackage{caption}
\usepackage{subfigure}
\usepackage{graphicx}
\usepackage{float}
\usepackage{booktabs}
\usepackage{setspace}
\usepackage{array}
\usepackage{makecell}
\usepackage{multirow}

\author[Ping Wang et al.]{{Ping Wang, Ping Chen, Zhimin Luo, Gaofeng Dong, Mengce Zheng, Nenghai Yu, Honggang Hu}}
\institute{Key Laboratory of Electromagnetic Space Information, CAS\\ University of Science and Technology of China, Hefei 230026, China \email{{wangping,cp2017,zmluo,dgf,mczheng}@mail.ustc.edu.cn, {hghu2005,ynh}@ustc.edu.cn}}

\title[Enhancing Practical Profiling Side-Channel Attacks Using CGAN]{Enhancing the Performance of Practical Profiling Side-Channel Attacks Using Conditional Generative Adversarial Networks}

\begin{document}

\maketitle

\keywords[Profiling side-channel attacks, CGAN, Generated traces, Leakage learning, Insufficient profiling set]{Profiling side-channel attacks \and CGAN \and Generated traces \and Leakage learning \and Insufficient profiling traces}

\begin{abstract}
    Recently, many profiling side-channel attacks based on Machine Learning and Deep Learning have been proposed. Most of them focus on reducing the number of traces required for successful attacks by optimizing the modeling algorithms. In previous work, relatively sufficient traces need to be used for training a model. However, in the practical profiling phase, it is difficult or impossible to collect sufficient traces due to the constraint of various resources. In this case, the performance of profiling attacks is inefficient even if proper modeling algorithms are used.
    
    In this paper, the main problem we consider is how to conduct more efficient profiling attacks when sufficient profiling traces cannot be obtained. To deal with this problem,  we first introduce the Conditional Generative Adversarial Network (CGAN) in the context of side-channel attacks. We show that CGAN can generate new traces to enlarge the size of the profiling set, which improves the performance of profiling attacks.
    For both unprotected and protected cryptographic algorithms, we find that CGAN can effectively learn the leakage of traces collected in their implementations. We also apply it to different modeling algorithms. In our experiments, the model constructed with the augmented profiling set can reduce the required attack traces by more than half, which means the generated traces can provide useful information as the real traces.
\end{abstract}


\section{Introduction}
Side-Channel Attacks (SCA) was first introduced by Paul Kocher \cite{DBLP:conf/crypto/Kocher96} in 1996, conducting a timing attack on the cryptographic device to recover the secret information. This provides a new direction for the security evaluation of the cryptographic algorithm when considering physical implementation. In addition to timing, the leakages available for SCA also include power consumption and electromagnetic emanations, etc. SCA mainly includes two categories: \emph{profiling} attacks and \emph{non-profiling} attacks. Non-profiling attacks mean that we recover secret information directly by analyzing the leakages of an unknown device and during the process, we need to guess the secret key. Typical non-profiling attacks include Differential Power Analysis (DPA) \cite{DBLP:conf/crypto/KocherJJ99}, Correlation Power Analysis (CPA) \cite{DBLP:conf/ches/BrierCO04}, and Mutual Information Analysis (MIA) \cite{DBLP:conf/ches/GierlichsBTP08}, etc. Profiling attacks (e.g. Template Attacks (TA) \cite{DBLP:conf/ches/ChariRR02}, and Stochastic Attacks (SA)\cite{DBLP:conf/ches/SchindlerLP05}) happen in the circumstance where the same fully-controlled profiling device exists before recovering the secret information used in a targeted device. Then we characterize the leakages of the controlled device by modeling algorithms and recover the secret information through the constructed model.

With the development of SCA, there appeared profiling attacks based on \emph{Machine Learning} (ML) \cite{DBLP:journals/jce/HospodarGMVV11,DBLP:conf/cardis/BartkewitzL12,DBLP:conf/cardis/LermanMBM13,DBLP:journals/ijact/LermanBM14}, in which two commonly used models are SVM \cite{DBLP:journals/ml/CortesV95,weston1998multi} and RF \cite{DBLP:journals/ml/Breiman01}. Recently, \emph{Deep Learning} (DL) based profiling attacks raises the concern of SCA community\cite{DBLP:conf/space/MaghrebiPP16,DBLP:conf/ches/CagliDP17,DBLP:journals/iacr/ProuffSBCD18,DBLP:journals/tches/KimPHBH19,DBLP:journals/tches/PicekHJBR19}. Compared to classical TA, DL-based profiling attack is no need for the Points of Interest (POI) or the traces alignment, and it can also handle high-dimensional data.
However, we find that these works focus on how to propose better modeling algorithms, or to improve the current modeling algorithms, in order to reduce the number of traces required for successful attacks. In fact, the traces they use in the profiling phase are relatively sufficient, which seems rational. Because the premise of profiling attacks is that the attackers have a fully-controlled profiling device and are able to obtain sufficient traces to construct the model. 

However, if the profiling attack is applied to a specific application, due to various constraints like time, resources, and countermeasures, we cannot collect enough traces. For example, an encryption operation to be attacked may only be a subroutine in the entire application, but the encryption operation has many pre-operations, so in the process of collecting traces, it is necessary to wait for the completion of the entire pre-operation. Even the position of the encryption operation in the entire application is not fixed. In this case, the trace of the entire process executed by the application needs to be collected, so each trace collection requires a huge time cost. Besides, in some application scenarios, the key has a life cycle. For example, the session key used when communicating between two devices, then the number of traces collected for each session key is limited, depending on how long the session time is. In such a scenario, the attack performance of the model constructed with insufficient traces is inefficient, and even an effective attack model cannot be constructed. Consequently, the attack performance of the model constructed by various approaches based on ML and DL mentioned above will be greatly decreased when the traces are lacking during the profiling phase.

Therefore, we consider whether it works by using the existing insufficient profiling traces to generate new traces with more useful information. If it is possible, we can use the generated traces to expand the original profiling set, to build a more efficient model. We find that Generative Adversarial Networks (GAN) \cite{DBLP:conf/nips/GoodfellowPMXWOCB14} proposed in 2014 has the ability to generate new data with additional information from raw data. GAN has a very rapid development in recent years and derives hundreds of different networks. Using GAN to generate data is mainly utilized in the field of image \cite{DBLP:journals/corr/abs-1711-04340,DBLP:conf/isbi/Frid-AdarKAGG18,DBLP:journals/corr/abs-1810-10863}. For example, Bowles \emph{et al.} used GAN to generate medical images \footnote{Because the annotation of medical images is expensive and time-consuming, it is impossible to get a lot training samples.}, which obtained satisfactory results\cite{DBLP:journals/corr/abs-1810-10863}. Thus, we try to use GAN to generate traces, to deal with the problem that sufficient profiling traces cannot be collected in the practical profiling attack. It should be noted that during the profiling phase, each trace has a corresponding label, thus we use Conditional Generative Adversarial Networks (CGAN) \cite{DBLP:journals/corr/MirzaO14} which can specify the labels of the generated traces. To the best of our knowledge, the method of using CGAN to generate traces has not yet been studied in the context of SCA. In our experiments, we first explore how to generate effective traces, and propose methods to evaluate the quality of generated traces. Our results show that CGAN can effectively learn the leakages of traces and generate traces with additional useful information, thereby improving the attack performance in scenarios where profiling traces are insufficient.

\subsection{Related Work}

The method of generating new traces then added to the original profiling set is also called Data Augmentation (DA) \cite{DBLP:conf/icdar/SimardSP03}. In recent years, the DA technique has been applied to profiling attacks. In 2017, Cagli \emph{et al.} mitigated overfitting and obtained better attack performance by randomly shifting traces, adding clock jitter, and then adding deformed traces to the original profiling set \cite{DBLP:conf/ches/CagliDP17}. In 2019, Picek \emph{et al.} used the Synthetic Minority Oversampling Technique (SMOTE) to process imbalanced data caused by using Hamming weight (HW) (or the Hamming distance (HD)) \cite{DBLP:journals/tches/PicekHJBR19}. At the same time, we notice that Picek \emph{et al.} proposed a restricted attacker framework where they restrict the ability of an attacker to obtain measurements in the profiling phase\cite{DBLP:journals/iacr/PicekHG19}. Although we both study similar issues,  the methods used are completely different.  In fact, the DA techniques previously mentioned improving attack performance are used in the scenario of relatively sufficient profiling traces. It should be noted that we concentrate on the problem of how to build a more efficient model when the collected traces are insufficient during the profiling phase. We use the limited profiling set to generate new traces with additional useful information to effectively improve the attack performance.

\subsection{Our Contributions}

Our contributions mainly include the following aspects:
\begin{enumerate}[leftmargin=\parindent, itemsep=0.1pt]
	\item We give a new perspective against the problem that how to build a more efficient model with insufficient traces in the practical profiling attacks. Then we first introduce CGAN (GAN) in the SCA context, and use it to generate traces to solve the problem of insufficient profiling traces, which improves the potential to apply profiling attacks in real scenarios.
	\item We present a detailed discussion on the label used to generated traces and evaluating the quality of generated traces. Our results show that the number of traces required for successful attacks can be reduced by more than half in the scenario of insufficient profiling set.
	\item For various types of AES implementations, such as unprotected, first-order masked protected, and random delay protected ones, we find that CGAN can effectively learn the leakages contained in traces.
	\item We have tried different models based on ML and DL, and find that our method is suitable for different modeling algorithms.
\end{enumerate}

\subsection{Outline}

The paper is organized as follows. In Section \ref{sec:background} we give an overview of profiling attacks, evaluation metrics, DA and CGAN (GAN). In Section \ref{sec:cgan_sca}, we describe the basic steps of using CGAN to generate traces and improve the attack performance and then verify the feasibility of the method through simulation experiments. In Section \ref{sec:exp}, through the experiments with real datasets, we verify that our method can improve the performance of profiling attacks in the case of insufficient profiling traces. The applicability of this method to different AES implementations and different modeling algorithms is also studied. In Section \ref{diss}, we give a discussion. Finally, in Section \ref{sec:conc} we have a summary of the proposed method and the future outlook.

\section{Background}
\label{sec:background}

\subsection{Profiling Side-Channel Attacks}
\label{subsec:psca}

We first briefly introduce the TA which was proposed in 2002 by Chari \emph{et al.} \cite{DBLP:conf/ches/ChariRR02} and is also the earliest profiling attack. TA consists of two phases: the profiling phase and the attack phase. During the profiling phase, it assumes that before the secret key of an unknown device is recovered, an identical device already exists and is fully controlled. The attackers calculate the intermediate values based on the plaintexts/ciphertexts and the fixed possible key and then divide the collected traces (or profiling set) according to the intermediate values. Finally, templates are established by using a Multivariate Gaussian distribution to characterize the relationship between the intermediate values and the leakages. In the attack phase, the attackers collect the traces of the unknown device and use the templates to recover the secret key used \footnote{We usually use a divide-and-conquer method to recover the subkey first.}. Subsequent profiling attacks based on ML and DL are basically the same as TA except that the modeling algorithms used are different.

Formally, during the profiling phase, the attackers get a profiling set $T_{profiling}=\{\vec{l_i}|i\le N_p\}$ with known plaintexts/ciphertexts $P_{profiling}=\{p_i|i\le N_p\}$ and the fixed key $k_p^*$. Function $\varphi$ computes the intermediate value $z_i$ between $p_i$ and $k_p^*$. Then, the model is constructed by the modeling algorithm $F$: 
\begin{equation}
M_z=F(\vec{l_i})\ where \ z=\varphi(p_i,k_p^*).
\end{equation}
$M_z$ is a model constructed with all the traces corresponding to the same intermediate value $z$, consequently we obtain models for each different intermediate value.

Similarly, we assume the attackers get attack set $T_{attack}=\{\vec{l_j}|j\le N_a\}$ with known plaintexts/ciphertexts $P_{attack}=\{p_j|j\le N_a\}$ during attack phase. Then we use \emph{Maximum Likelihood} strategy to calculate the score $S$ for each guessed key $k_{guess}$: 
\begin{equation}
S_{k_{guess}}=\prod_{j=1}^{N_a}M_{\varphi(p_j,k_{guess})}(\vec{l_j}).
\end{equation}
The $k_{guess}$ that maximizes $S$ is the key that the model thinks correct: 
\begin{equation}
k_{real}=\mathop{\arg\max}_{k_{guess}}S_{k_{guess}}.
\end{equation}
Note that, ML and DL-based profiling attacks usually build only one model, but will output scores (or probability) corresponding to all intermediate values in the attack phase, which is equivalent to a model including all models $M_z$.

\subsection{Evaluation Metric}

In ML and DL, accuracy is often used as an evaluation metric. Although we use the ML and DL based methods in profiling attacks, some researchers have found that using accuracy to evaluate SCA is not suitable \cite{DBLP:journals/tches/PicekHJBR19}. Hence, we will use Guessing Entropy (GE) \cite{DBLP:conf/eurocrypt/StandaertMY09}, one of the common SCA metrics. In simple terms, GE is to calculate the averaged rank of the correct key over several experiments, and the term rank refers to the score ranking of all key hypotheses given by the model. What we usually do is to calculate the average number of traces required when the rank converges to zero. In our experiments, after CGAN learns the distribution of the original data, the generated traces each time are different and the improvement  of the attack result also changes slightly. Therefore, in each experiment, we will generate new traces and the original training set and testing set are randomly selected. Then we average the attack results of 10 experiments as GE.

\subsection{Data Augmentation}

Data Augmentation \cite{DBLP:conf/icdar/SimardSP03} is a regularization technique used to deal with the overfitting problem, in order to build a more robust Model. Naturally, this method is often used in the field of image. For example, Krizhevsky \emph{et al.} used two forms of DA techniques to effectively reduce overfitting, one of which is image translations and horizontal reflections, and the other is altering the intensities of the RGB channel \cite{DBLP:conf/nips/KrizhevskySH12}. Since GAN was introduced in 2014, some GAN-based Data Augmentation has been proposed and used to acquire good results \cite{DBLP:journals/corr/abs-1711-04340,DBLP:conf/isbi/Frid-AdarKAGG18,DBLP:journals/corr/abs-1810-10863}. In the field of SCA, there have also been some researches \cite{DBLP:conf/ches/CagliDP17,DBLP:conf/cardis/PuYWGLGWG17,DBLP:journals/tches/PicekHJBR19} on the use of DA technique recently.
Besides, it should be emphasized that our main concern is how to build a more efficient model when sufficient traces cannot be collected during the profiling phase, instead of reducing overfitting \footnote{In fact, when we use CGAN to generate traces and expand the original profiling set, the overfitting problem will also be alleviated, but this is not in conflict with the main problem we are aiming at.}.

\subsection{Conditional Generative Adversarial Networks}
\begin{figure}[!t]
	\centering
	\includegraphics[width=.65\textwidth]{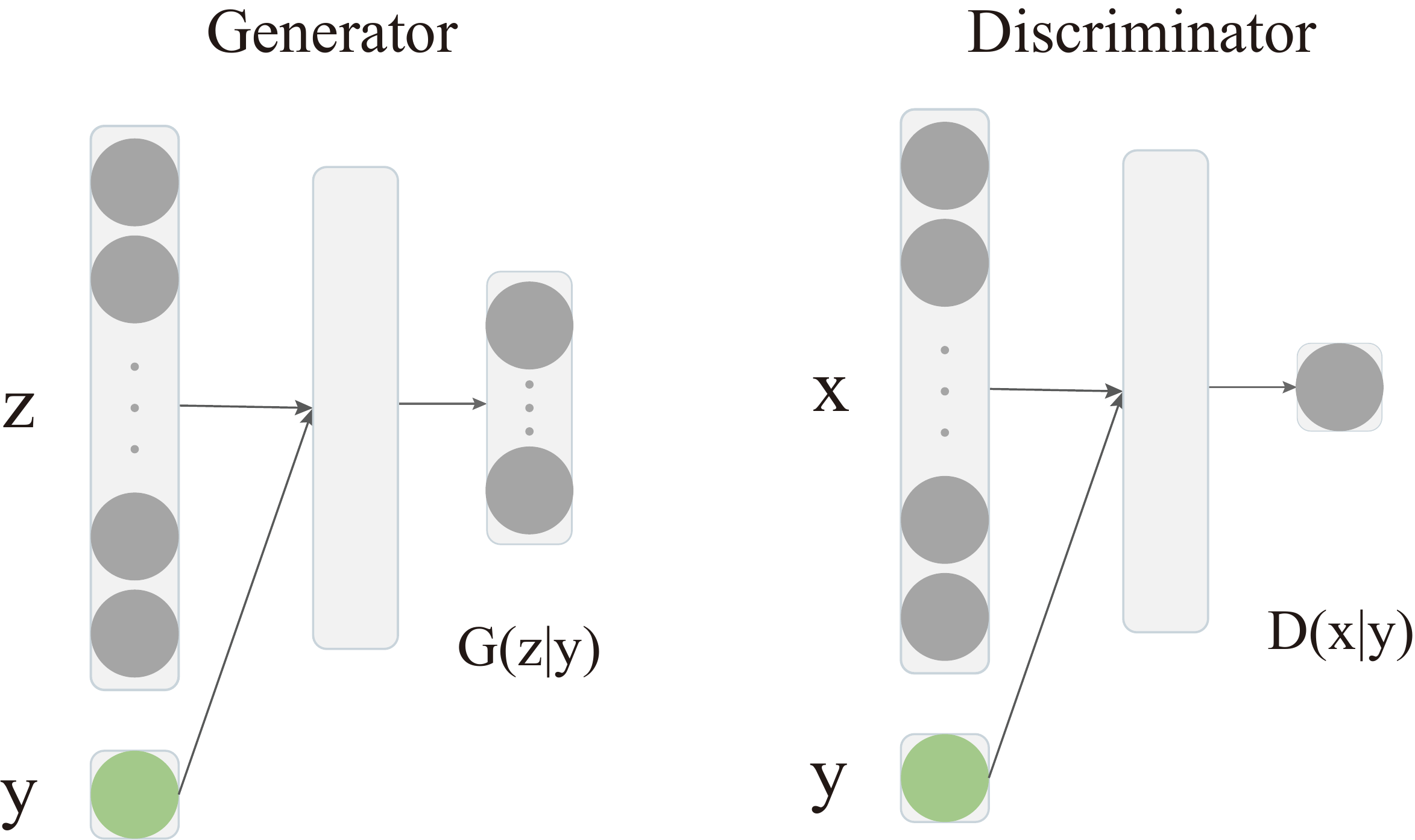}
	\caption{Conditional Generative Adversarial Network.}
	\label{fig1}
\end{figure}

In this section, we mainly introduce the basic principles of Generative Adversarial Networks (GAN) and Conditional Generative Adversarial Networks (CGAN). GAN \cite{DBLP:conf/nips/GoodfellowPMXWOCB14} was proposed in 2014 and mainly used as a generative model to learn the distribution of data. GAN includes two parts, generator $G$ and discriminator $D$. Assuming that the real data is $x$ and its distribution satisfies $p_{data}$, the role of the generator is to learn the distribution of real data $p_{data}$ through a non-linear function $G(z;\theta_g)$ which maps a prior noise distribution $p_z(z)$ to $p_{data}$. Assuming that the distribution of the learned data through the generator is $p_g$, all we have to do is to make $p_g$ as close to $p_{data}$ as possible. The discriminator is also a non-linear function $D(x,\theta_d)$, giving the probability that a sample comes from $p_{data}$ rather than $p_g$. Then we alternately train the discriminator and generator, and adjust the parameters $\theta_d$ and $\theta_g$ to maximize $\log{D(X)}$ and minimize $\log(1-D(G(Z)))$. This training process can be regarded as a mini-max game between the generator $G$ and the discriminator $D$. Using an objective function to express it is: 
\begin{equation}
\mathop{\min}_{G}\mathop{\max}_{D}V(D,G)= \mathbb{E}_{x\sim p_{data}(x)} \left[\log{D(x)} \right]+\mathbb{E}_{z\sim p_z(z)} \left[\log{(1-D(G(z)))}\right].
\end{equation}
GAN is actually a kind of unsupervised learning, and the generated data has no label. However, the profiling SCA based on ML and DL is supervised learning which means every trace has a corresponding label obtained through the intermediate value and energy model. Therefore, each generated trace must also have a corresponding label so that the original profiling set can be expanded. Consequently, we use a conditioned version of GAN, namely CGAN \cite{DBLP:journals/corr/MirzaO14}, to generate traces. The difference between CGAN and GAN is that in CGAN when training $G$ and $D$, we add additional information (or auxiliary information) $y$, which can be the label of the data or the data under a certain distribution. Then, the target function becomes:
\begin{equation}
\mathop{\min}_{G}\mathop{\max}_{D}V(D,G)= \mathbb{E}_{x\sim p_{data}(x)} \left[\log{D(x|y)} \right]+\mathbb{E}_{z\sim p_z(z)} \left[\log{(1-D(G(z|y)))}\right].
\end{equation}
Figure \ref{fig1} simply shows the architecture of CGAN. After using CGAN, we can specify label and produce the corresponding traces.

\section{CGAN on SCA}
\label{sec:cgan_sca}
In this section, we outline the overall process of how to use CGAN to enhance the performance of profiling attacks when the profiling traces are insufficient and verify the feasibility of this method by simulation experiments. Then, we discuss some experimental parameters that need attention when traces are generated.
\subsection{The Basic Steps}

\subsubsection{Choosing an Insufficient Training Set}
\label{sec:smallset}
First, we divide the traces into two parts, a set of training traces and a set of testing traces, respectively corresponding to the profiling phase and the attack phase. Afterward, we reduce the size of training set as much as possible to simulate the scenario of insufficient profiling traces. In our experiments, we use the training sets of different sizes to build models in advance and compare the attack performance on the same testing set. Then we select a training set with general attack performance whose set size is relatively small \footnote{We apply the same strategy in subsequent experiments, i.e., to choose the size of the training set beforehand and use a relatively small training set to simulate the lack of traces during the profiling phase.}. In fact, when adequate traces cannot be collected during the profiling phase, all traces would be utilized maximally to construct the model.

\subsubsection{The CGAN Structure}

For the generator and discriminator, it is mainly composed of fully connected layers, and no more complex layer is used. However, in the process of CGAN training, instability is probable to occur. Here we use some techniques: (1) Use Batch Normalization (BN) in both Generator and Discriminator \cite{DBLP:journals/corr/RadfordMC15}, (2) The prior noise is sampled from the Gaussian distribution rather than Uniform distribution \cite{DBLP:journals/corr/White16a}, (3) Use the leaky rectified activation (LeakyReLU) \cite{maas2013rectifier} as the activation function for the hidden layer, (4) Use the Adam optimizer \cite{DBLP:journals/corr/KingmaB14} to train the model, (5) Use the Dropout layer in generator during both training and testing phases \cite{DBLP:conf/cvpr/IsolaZZE17}. In addition, tuning CGAN hyperparameters is similar to tuning classifiers. The difference is that the optimal parameters should be selected according to the performance of the model constructed with the augmented training set. At the same time, it should be noted that we should use the evaluation metrics related to side-channel attacks. The specific model structure of CGAN is given in Appendix \ref{appendix_cgan}.

\subsubsection{Evaluating the Generated Traces}
\label{sec:evaluate}
After traces are generated, there arises a question of how we evaluate the quality of the generated traces. When using the technology related to GAN to generate images, there are many common evaluation metrics, such as Accuracy, IS \cite{DBLP:conf/nips/SalimansGZCRCC16}, FID \cite{DBLP:conf/nips/HeuselRUNH17}. However, these metrics are not necessarily applicable for SCA. For example, paper \cite{DBLP:journals/tches/PicekHJBR19} verified that ML evaluation metrics are not precise for sure in SCA evaluations, such as accuracy. Even if the generated traces are similar to the real traces, we cannot decide whether the generated traces can contribute to the model. Therefore, in the final analysis, we need to add the generated traces to the original profiling set and see if the attack results are improved, in order that we can determine whether the generated traces are effective. That is to say, the quality of the generated traces comes down to the extent of improvement they bring to the attack performance.

Additionally, there are other intuitive methods to estimate the quality of generated traces. One of them is to visually observe the shapes of generated traces. If the generated traces seem largely different from the original traces, such generated traces are not expected to yield better improvement. This is because the purpose of GAN is exactly to learn the distribution of the original data and the generated objects should be similar to the original objects. Another method is to use Correlation Power Analysis (CPA) \cite{DBLP:conf/ches/BrierCO04} to separately calculate the correlation of original traces and generated traces, then compare the location of the leaks and the intensity of the correlation to verify whether the leakages have been learned. The paper \cite{DBLP:journals/tches/Timon19} also mentioned that the leakage points make the greatest contribution to modeling, thus observing the correlation of the generated traces at the leakage points is the simplest and most effective way to assess the quality of the generated traces. Based on multiple experiments, we discover that if the CPA results between the generated traces and the original traces are closer, the enhancement of the attack performance after data expansion is more obvious. 

We also tried to use DPA \cite{DBLP:conf/crypto/KocherJJ99} to evaluate whether the generated traces have learned effective leakage, and the experimental results show that it is indeed feasible. See Appendix \ref{dpa} for the detailed results of using DPA to evaluate the generated traces.

\subsubsection{The Way to Improve the Attack Performance}
\label{sec:twoway}

We must emphasize that we are talking about the same model, that is, the ways are on the basis of the premise that the overall structure of the model is unchanged \footnote{Our goal is not to find the optimal model, but to figure out whether using generated traces to expand original training set in a scenario where profiling traces are insufficient can improve the attack performance. Therefore, we ought to assume that the model structure remains the same.}. Without considering various pre-processing methods, then the size of the profiling set used to construct the model determines the performance of the attack.

The idea of the method is simple, we can utilize the existing insufficient profiling traces to generate new traces and expand original profiling set to obtain better attack performance. Let $T_{raw}$  be the original profiling set, $M_{raw}$ be the model built on $T_{raw}$, $P(M_{raw})$ be the attack performance of the model $M_{raw}$. Let $T_{generated}$ be the traces generated by $T_{raw}$. Then we can build model $M_{raw+generated}$ with  $T_{raw}$ and $T_{generated}$. Finally, if the generated trace is effective, we can get $P(T_{raw}+T_{generated})\textgreater P(T_{raw})$.

\subsection{Simulation Experiments}
\label{sec:sim}

Firstly, we generate $N=10000$ simulated traces \footnote{The method of simulated traces generation is referred to in the paper \cite{DBLP:journals/tches/Timon19}.}, each of which has 50 time samples, and is related to the fixed key $k^*$. The plaintext $p_i$ related to each trace is randomly generated between $0\sim255$ and the amplitude of each time sample is also a randomly value between $0\sim255$. The leakage point is located at $t_{leakage}=25$, and the amplitude of the leakage point is $Sbox(k^*\oplus p_i)+\mathcal{N}(0,1)$.

In the experiment, two disjoint sets are randomly divided out from the first 5000 simulated traces, the training set with 500 traces and the validation set with 2000 traces. Besides, 2000 traces are divided out from the last 5000 simulated traces as the testing set.

Secondly, we build an MLP model with two hidden layers and the numbers of neurons in the two layers are separately 70, 50 and the label is Least Significant Bit (LSB) of $Sbox(k^*\oplus p_i)$. Following the steps in \autoref{sec:smallset}, we choose the size of the training set to be 500 to simulate the scenario where the profiling traces are insufficient.

\begin{figure}[!t]
	\centering
	\begin{minipage}[c]{0.48\textwidth}
		\centering
		\includegraphics[width=7.5cm]{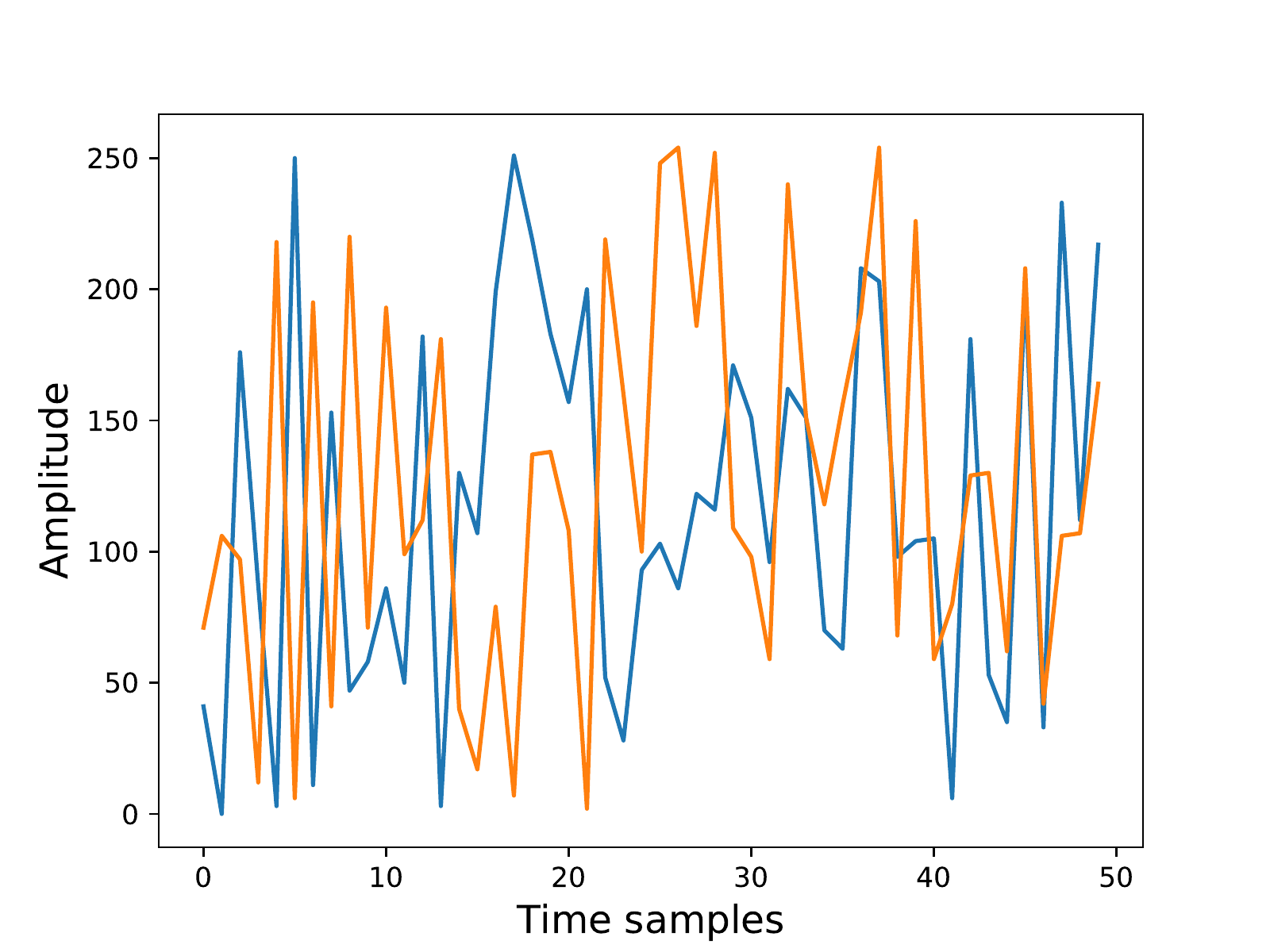}
	\end{minipage}
	\hspace{0.01\textwidth}
	\begin{minipage}[c]{0.48\textwidth}
		\centering
		\includegraphics[width=7.5cm]{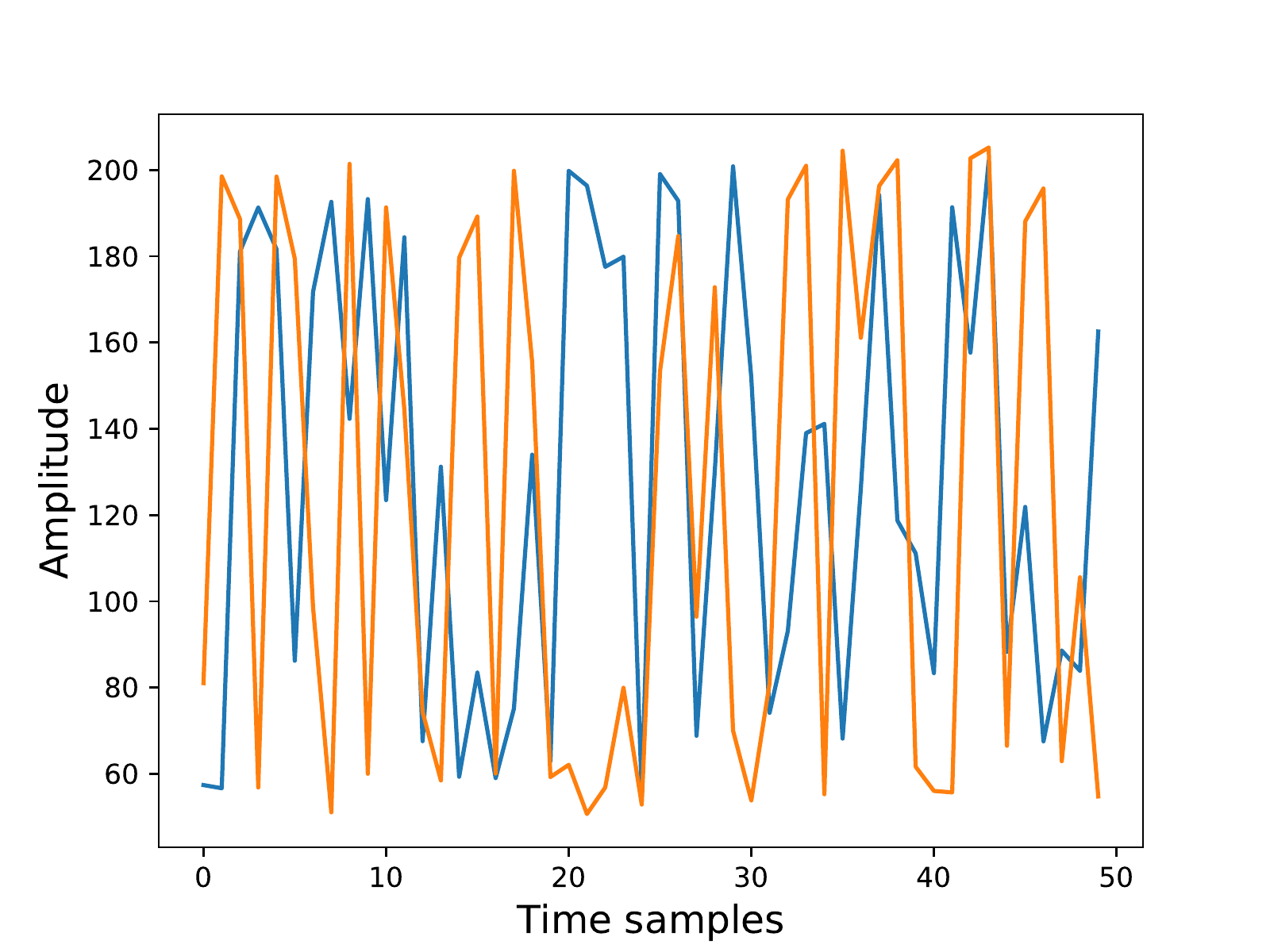}
	\end{minipage}
	\caption{Left: original traces. Right: generated traces.}
	\label{fig2}
\end{figure}

Thirdly, we use the 500 traces of the original training set to generate 400 traces, among which 200 traces are with label 0 and another 200 traces are with label 1 \footnote{In the later experiments, we will specify that the number of generated traces with labels 0 and 1 are equal to each other when we use LSB as the label.}. We randomly choose two traces from the original training set and the generated traces separately and they are shown in Figure \ref{fig2}.

\begin{figure}[!t]
	\centering
	\begin{minipage}[c]{0.48\textwidth}
		\centering
		\includegraphics[width=7.5cm]{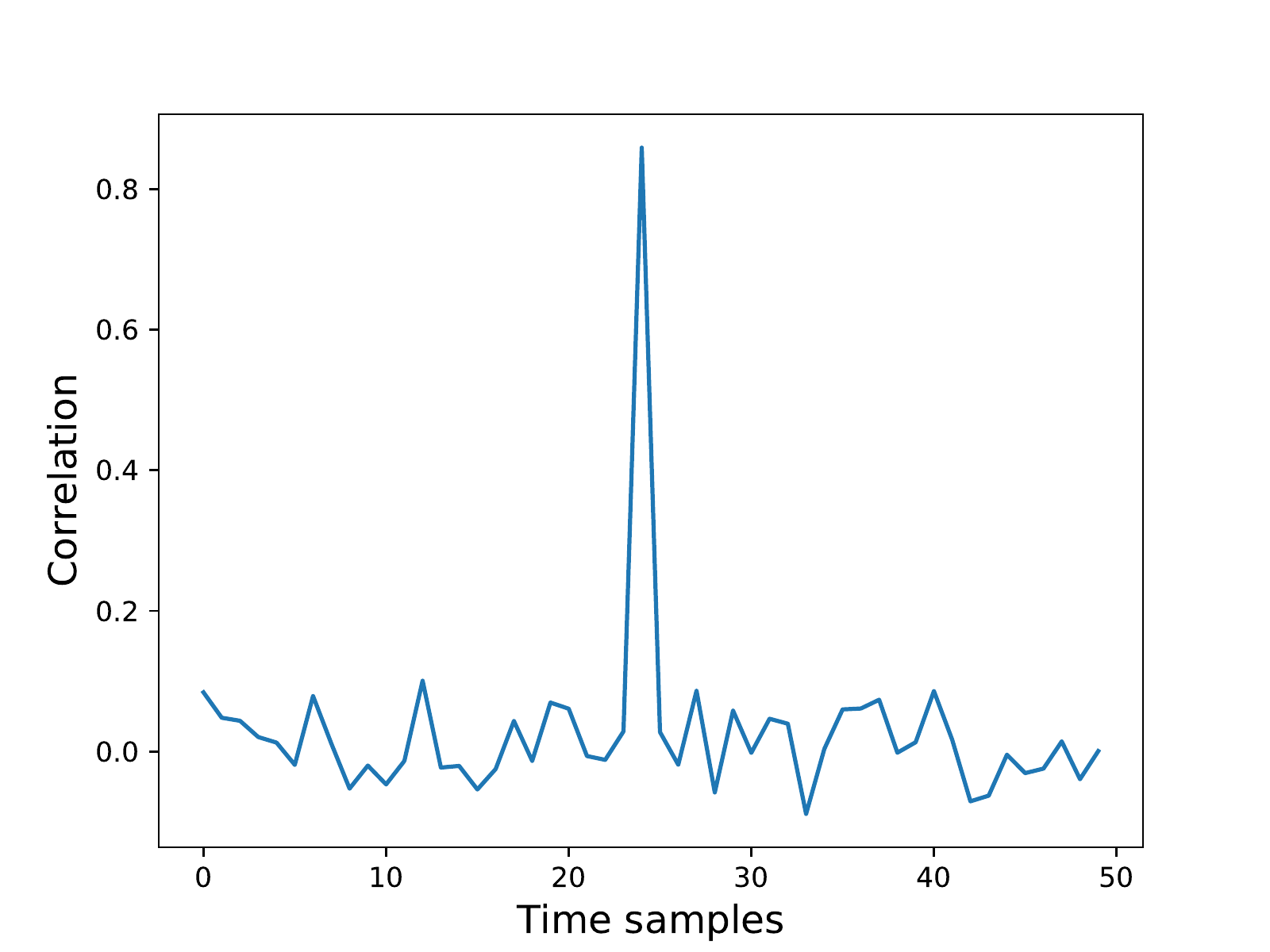}
	\end{minipage}
	\hspace{0.01\textwidth}
	\begin{minipage}[c]{0.48\textwidth}
		\centering
		\includegraphics[width=7.5cm]{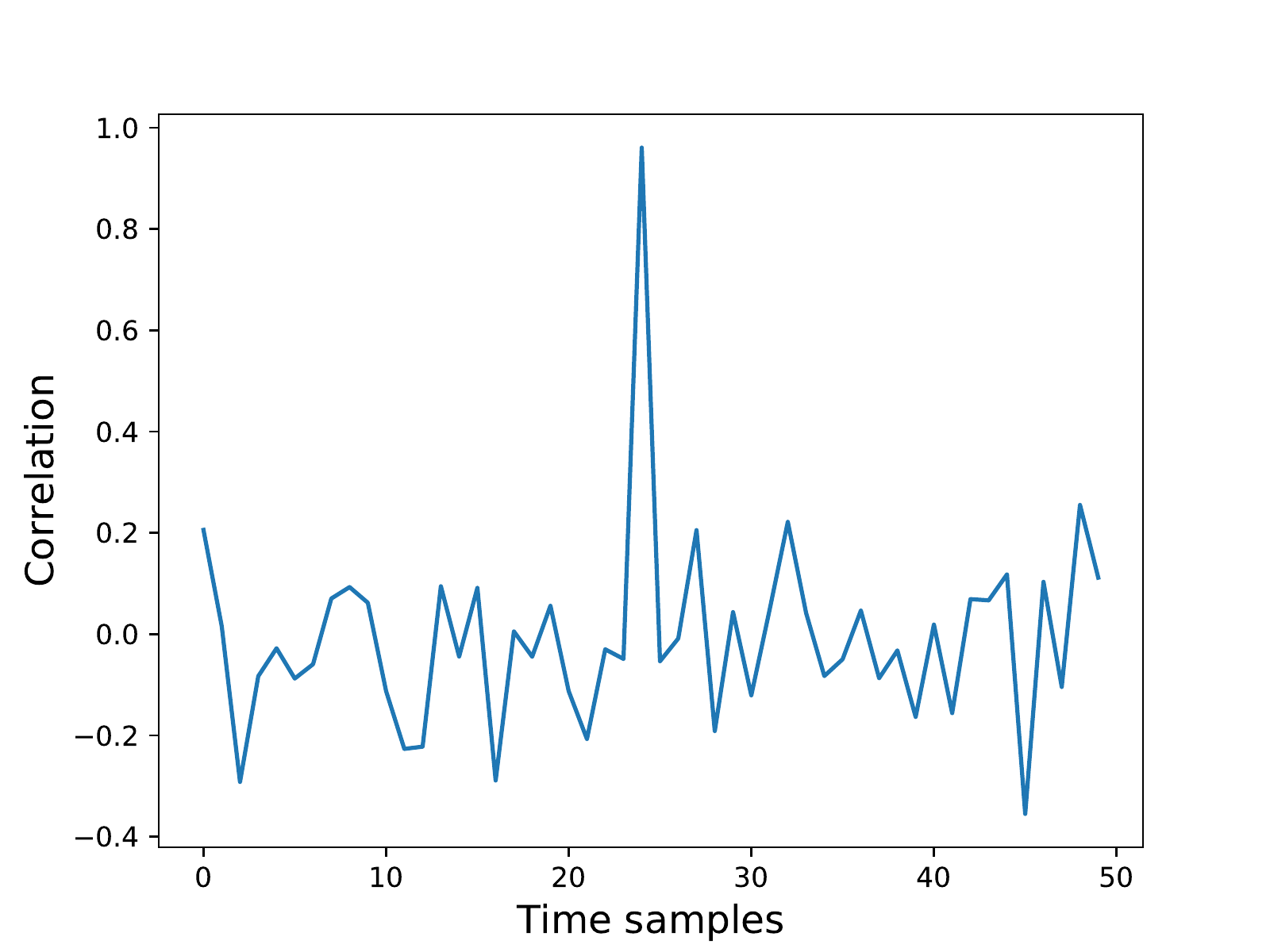}
	\end{minipage}
	\caption{Left: CPA of 500 original traces. Right: CPA of 400 generated traces.}
	\label{fig3}
\end{figure}

Then, we continue to analyze the correlation of the 500 original training traces and the 400 generated traces, and compare the CPA results in these two cases and determine whether the leakages have been learned through the currently generated traces. The CPA results are illustrated in Figure \ref{fig3}.	

It indicates that in the CPA using the generated traces, the location of the leakage point and the amplitude of leakage peak are consistent with the CPA using the original training set. Meanwhile, we also see that the correlation of other time samples is higher in the CPA using the generated traces, and we call it \emph{leakage noise}. The reason for the leakage noise may be that when CGAN learns the distribution of the original traces, the generator would think that it is easier to fool the discriminator when the leakages appear at more locations. In the later experiments, we will prove that the generated traces with high leakage noise are unable to improve the performance of the modeling attack, or even reduce the performance.

Fourthly, as we point out in \autoref{sec:evaluate}, although the CPA results of the generated traces are close to that of the original traces, applying such generated traces for trace augmentation does not mean that the attack performance will definitely be improved. As a consequence, we need to decide whether our method is feasible by the practical attack results. For comparison, we randomly select 400 traces from the original 500 traces and add the 400 traces to the original training set to confirm that the addition of generated traces does improve the attack performance. The specific steps of the experiment are consistent with \autoref{sec:expsetup}, which we will not repeat here. The attack results are presented in Figure \ref{fig4}. 

It is apparently seen that using the augmented training set to build the model can achieve better attack performance, where the number of traces required for GE to converge to 0 is reduced by about 400. The results reveal that the generated traces are capable of providing additional useful information, which is equivalent to the circumstance where we collect more traces. Of course, the generated traces would make less contribution due to leakage noise in most cases.


\noindent\begin{minipage}[t]{.48\textwidth}
	\centering
	\includegraphics[width=7.5cm]{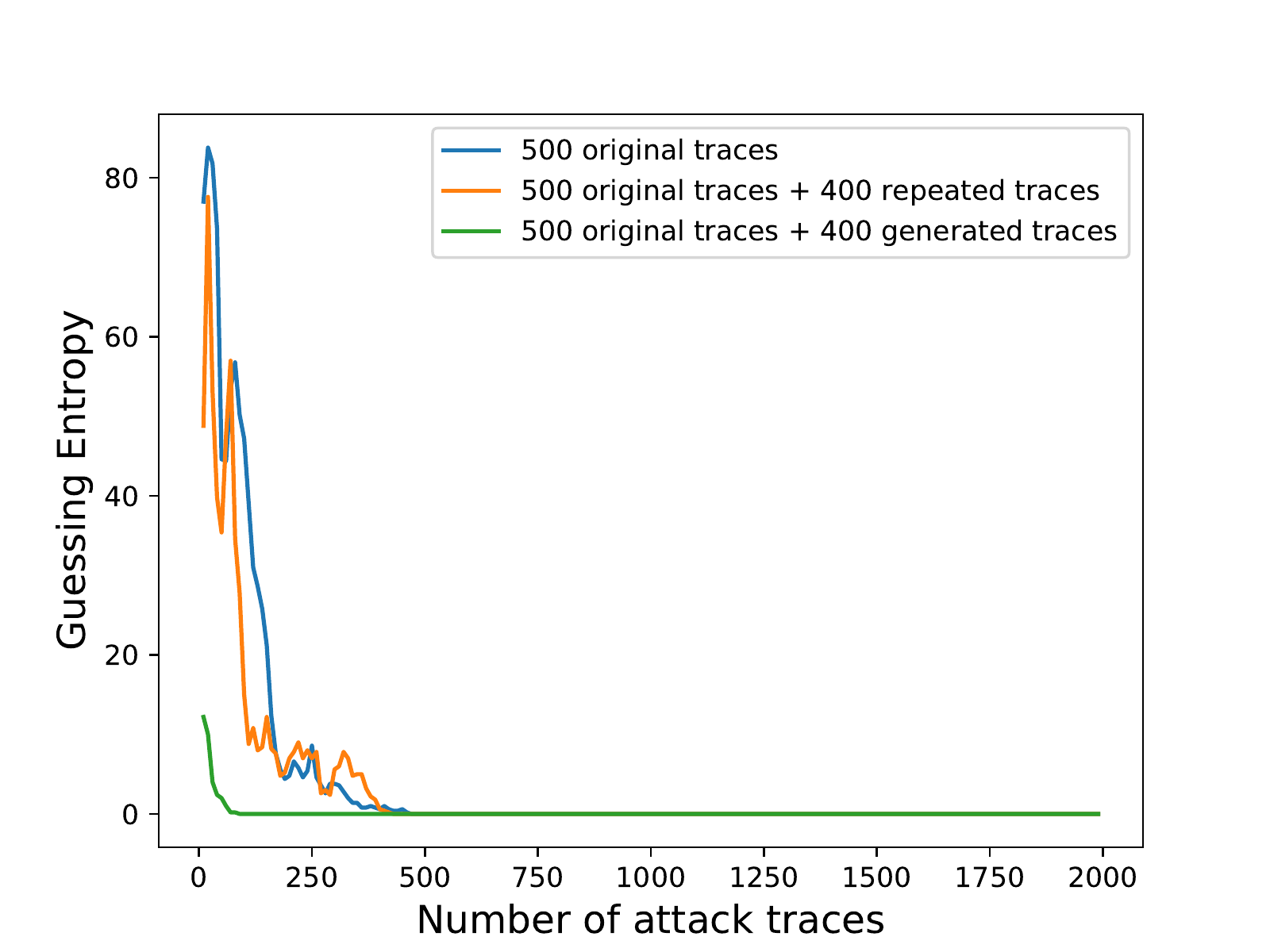}
	\captionof{figure}{The attack results of models built with \{500 original traces, 500 original traces + 400 repeated traces, 500 original traces + 400 generated traces\}.}
	\label{fig4}            
\end{minipage}%
\hspace{0.02\textwidth}
\begin{minipage}[t]{.48\textwidth}
	\centering
	\includegraphics[width=7.5cm]{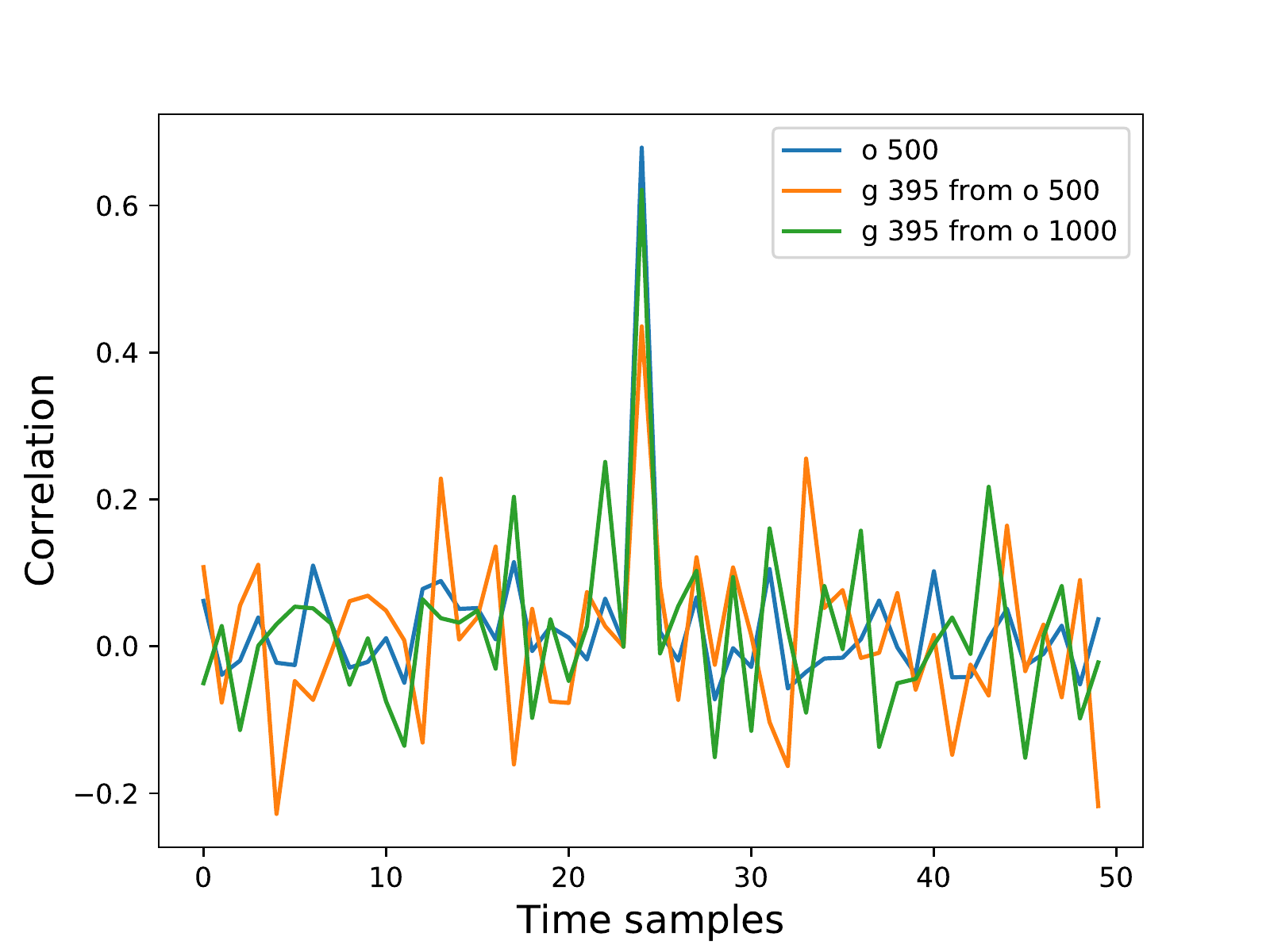}
	\captionof{figure}{CPA of \{500 original traces, 395 generated traces from 500 original traces, 395 generated traces from 1000 original traces\} when using HW as label.\vspace{24pt}}
	\label{fig5}            
\end{minipage}

\subsection{Labels}
\label{hw}
We continue to discuss how to select the label in the process of generating traces. In \autoref{sec:sim} we have been using the label LSB, next we will apply the same experimental configuration, except that the label is replaced with HW. Note that for the label HW, there are 9 categories in total and they satisfy the binomial distribution. Thus, we generate 395 traces according to the distribution, among which the amount of each category conforms to the binomial distribution. Furthermore, for better comparison, based on the original training set with 500 traces, we add another 500 traces and then use the 1000 traces to generate traces. Figure \ref{fig5} shows the comparison of CPA results.

From Figure \ref{fig5} we can see that if only 500 traces are used to generate traces, the correlation coefficient at the leakage points calculated by the generated traces is reduced by about 0.2, and the noise of other non-leakage points is more obvious as well. Then we add three sets of generated traces to the original training set of 500 traces to construct the models. The attack results are presented in Figure \ref{fig6}.


\noindent\begin{minipage}[t]{.48\textwidth}
	\centering
	\includegraphics[width=7.5cm]{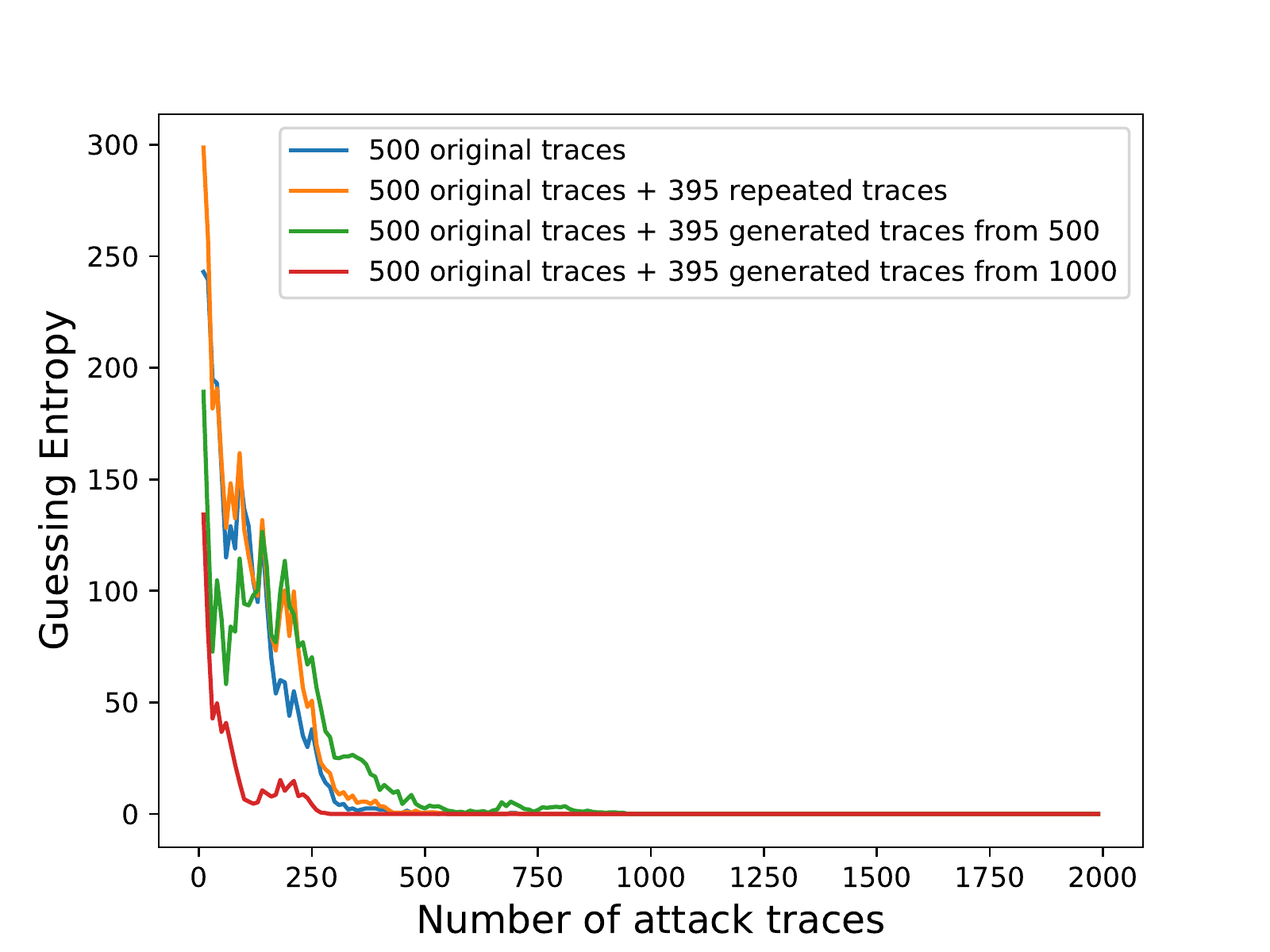}
	\captionof{figure}{The attack results of models built with \{500 original traces, 500 original traces + 395 repeated traces, 500 original traces + 395 generated traces from 500 original traces, 500 original traces + 395 generated traces from 1000 original traces\} when using HW as label.\vspace{24pt}}
	\label{fig6}            
\end{minipage}%
\hspace{0.02\textwidth}
\begin{minipage}[t]{.48\textwidth}
	\centering
	\includegraphics[width=7.5cm]{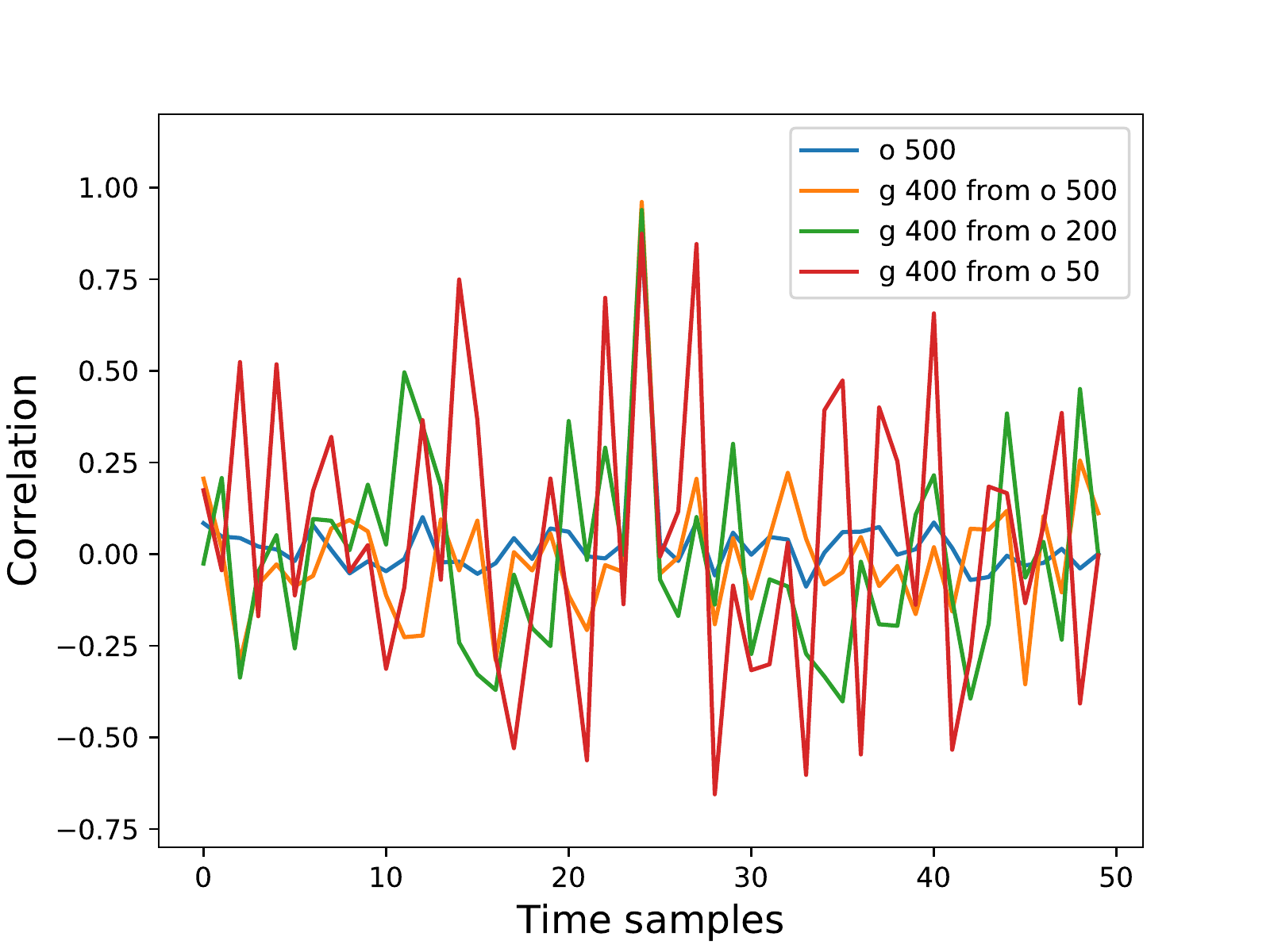}
	\captionof{figure}{CPA of \{500 original traces, 400 generated traces from 500 original traces, 400 generated traces from 200 original traces, 400 generated traces from 50 original traces\}.}
	\label{fig7}            
\end{minipage}

We find that the generated traces by using the 500 original traces do not bring about better attack performance, and even cause the attack result to become worse. We note that when the label HW is used, there exists a problem called imbalanced data \cite{DBLP:journals/tches/PicekHJBR19}. Similarly, when using the label HW for trace generation, we may also be confronted with this problem. Imagine that using a training set with 500 traces to generate traces, the amount of the traces with HW = 0, 8 maybe only 1 or 2, which makes it difficult for us to learn these two categories of traces. Therefore, the addition of these two categories of corresponding generation traces will not contribute to the model but aggravate the problem of data imbalance. While using 1000 traces for trace expansion produces slight enhancement, which indicates that using the label HW to generate traces is relatively difficult in the situation where the amount of existing original traces is small.

\subsection{Sizes of Training Set and Generated Trace Set}

In this section, we use training sets of different sizes to generate traces and study how the amount of traces in the original training set affects the quality of generated traces. We respectively use the original training sets of size 500, 200, and 50 to generate 400 traces with label LSB. Figure \ref{fig7} demonstrates the CPA results of generated traces using distinct training sets.

The CPA results show that though the correlation at the leakage points is similar, the noise at the non-leakage points increases as the size of the original training set is reduced. Especially, when only 50 traces are used for trace generation, the noise is strongly obvious, which tells us that if the original training traces are quite fewer, the generated traces would possess poor quality. Then, we add the generated traces above to the same original training set with 500  traces for modeling and observe whether the attack performance has been improved. The attack results are shown in Figure \ref{fig8}.


\noindent\begin{minipage}[t]{.48\textwidth}
	\centering
	\includegraphics[width=7.5cm]{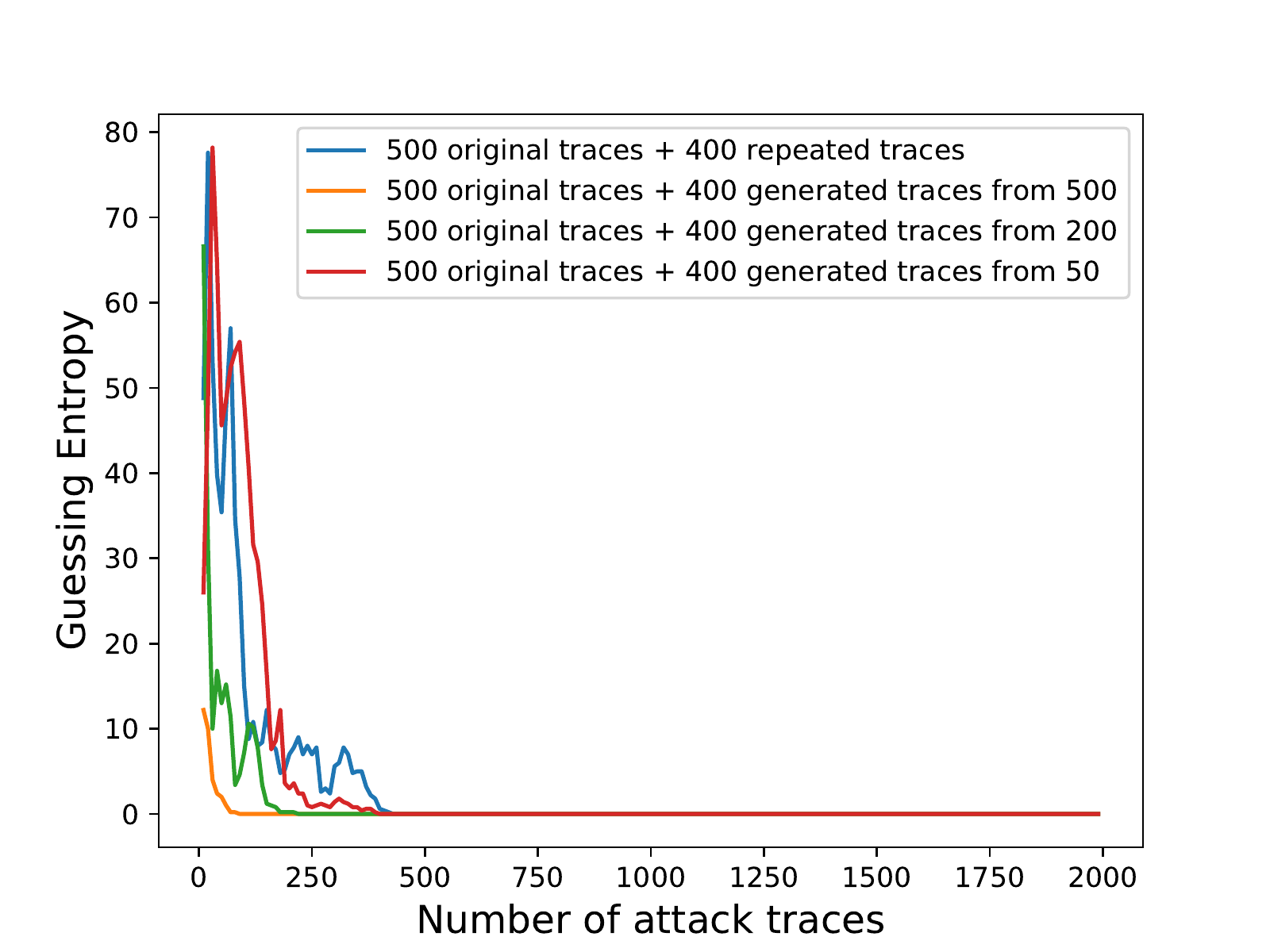}
	\captionof{figure}{The attack results of models built with \{500 original traces + 400 repeated traces, 500 original traces + 400 generated traces from 500 original traces, 500 original traces + 400 generated traces from 200 original traces, 500 original traces + 400 generated traces from 50 original traces\}.\vspace{24pt}}
	\label{fig8}            
\end{minipage}
\hspace{0.02\textwidth}
\begin{minipage}[t]{.48\textwidth}
	\centering
	\includegraphics[width=7.5cm]{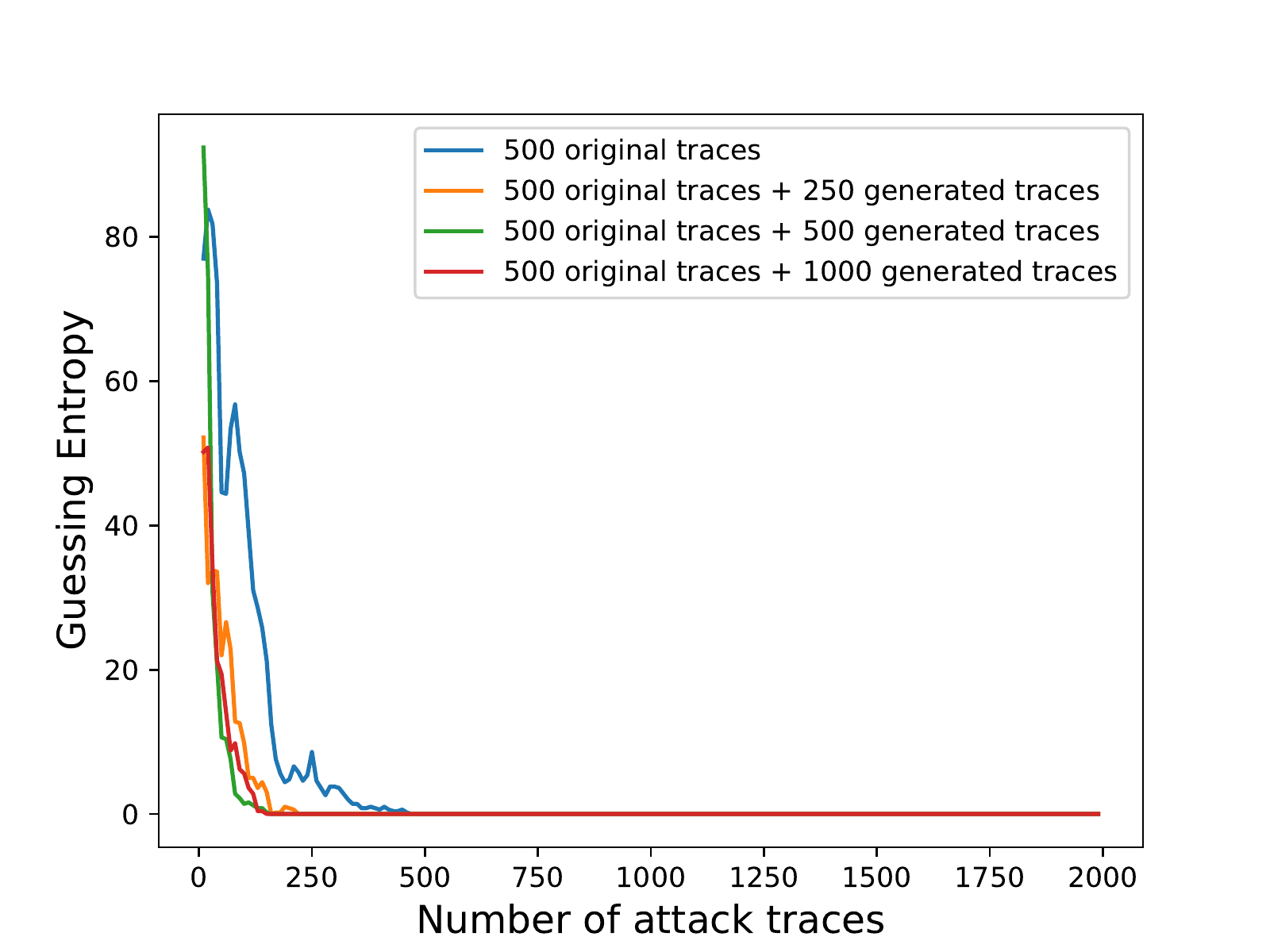}
	\captionof{figure}{The attack results of models built with \{500 original traces + 250 generated traces, 500 original traces + 500 generated traces, 500 original traces + 1000 generated traces\}.}
	\label{fig8}            
\end{minipage}
It implies that as the size of the original training set is decreasing, the improvement of the attack performance brought by the generated traces is constantly weakening. For example, when we only use 50 traces for trace generation and expansion, it is almost impossible to produce any amelioration. In the light of the above results and analysis, for effective generated traces, not only high correlation at the leakage points is required, but also low correlation at the non-leakage points is needed. In other words, we want the CPA results of the generated traces and the original traces to be as close as possible.

In addition to observing the effect of the size of original training set on the quality of the generated traces, we also have studied how the ratio between the size of added generated trace set and the size of original training set influence the attack performance. On the basis of previous experimental configuration, we use the 500 original traces to respectively generate 250, 500, and 1000 traces as three different generated trace sets, then add them to the 500 original traces in turn for modeling and compare the influence on attacks. The results are displayed in the following Figure \ref{fig9}.

The comparison of the attack results indicates that the ratio between the size of added trace set and the size of original training set does not actually affect the attack performance greatly. Hence, we are not going to discuss this issue in detail subsequently when we conduct experiments with real traces. Undoubtedly, if adding too few traces, it is hard to make any improvements. On the other hand, adding too many generated traces would make the model learn more from the generated traces, which will also reduce the attack performance.

\section{Experiments}
\label{sec:exp}

In this section, we first discuss the application of using CGAN to generate traces under three AES implementations: unprotected, first-order masked protected and random delay protected. Then we test the universality of the proposed method to different modeling algorithms. Recently, profiling attacks based on DL are becoming very popular, thus we mainly test common modeling algorithms used in DL.
\subsection{Experiment Setup}
\label{sec:expsetup}

In \autoref{sec:unprotected}, \autoref{sec:mask} and \autoref{sec:desync}, we will use the public model $MLP_{best}$ in \cite{DBLP:journals/iacr/ProuffSBCD18}. The reason why we use it is that our goal is to verify whether adding the generated traces to the profiling set can enhance the attack performance, rather than to find out the optimal model. Hence, we do not change the structure of the model, which has four hidden layers with 200 neurons in each layer. For different implementations of AES, we just modify the learning rate, epoch and batch size. Besides, we think the experimental results are more convincing by using public models.

Specifically, we first divide the data set into two parts $D_1$ and $D_2$. Then we randomly divide out the training set $D_{training}$ and the validation set $D_{validation}$ from $D_1$ in each experiment \footnote{$D_{training}$ and $D_{validation}$ are not intersected.}, and divide out the testing set $D_{testing}$ from $D_2$. Because of the overall structure of the model is fixed, we only need to fine-tune the learning rate, epoch, and batch size through $D_{validation}$ to get the best model $M_{original}$ \footnote{In fact, the learning rate, epoch and batch size corresponding to $M_{original}$ in each experiment will be basically unchanged. We do this because we consider that the size of the original training set would change after adding the generated traces, and the optimal value of the three parameters might also change, then we add a cross-validation process.} under the original training set $D_{training}$. Secondly, we use CGAN to generate trace set $D_{generated}$ through $D_{training}$ and combine $D_{generated}$ and $D_{training}$ as the new training set $D_{new\_training}$. We then use $D_{validation}$ to fine-tune the above parameters to get the best model $M_{new}$ under $D_{new\_training}$. Finally, we attack testing set $D_{testing}$ by using $M_{original}$ and $M_{new}$ respectively and get the score ranking of the correct key. We repeat the above experiment 10 times, with the average attack results as the final GE. Moreover, it must be emphasized that in order to make the final results more credible the number of traces in $D_1$ is greater than the sum of the number of traces in $D_{training}$ and $D_{validation}$, and the number of traces in $D_2$ is also greater than the number of traces in $D_{testing}$ in each experiment. 

To better verify the effectiveness of the traces generated by CGAN, we added three sets of comparative experiments for each data set. (1) add the repeated traces to the training set, (2) use the method of adding noise to the traces to obtain new traces and add them to the training set\cite{DBLP:journals/tches/KimPHBH19}, (3) add real traces to the training set. Another point that needs to be added is the comparison of the shape of the generated trace and the original trace, which we have placed in Appendix \ref{shape}.

\subsection{Unprotected AES}
\label{sec:unprotected}

In this section, we use two trace sets collected in two implementations of unprotected AES, one of which is DPAcontest v4 dataset, and the other is power trace set in unprotected AES-128 implementation running on an Atmel XMEGA128 chip, measured by ChipWhisperer-Lite (CW) platform \cite{CW}. In the following two subsections we verify that using generated traces can improve the effectiveness of profiling attacks with insufficient profiling traces. At the same time, we have found an interesting phenomenon that we can successfully perform profiling attacks only using generated traces.

\subsubsection{DPAcontest v4}
\label{sec:dpav4}
DPAcontest v4 dataset provides measurements in software implementation of AES-256 with Rotating Sbox Masking (RSM) protection \cite{DBLP:conf/date/NassarSGD12}. We can directly compute the masked S-box output through the known masks, thus turning it into an unprotected AES implementation. Here we only use 10,000 traces in the DPAcontest v4 dataset. We let the first 6,000 traces as $D_1$ and the remaining 4,000 traces as $D_2$. Then we randomly divide out 1,000 traces from $D_1$ as the training set and 2,000 traces as the validation set. We randomly choose 2,000 traces from $D_2$ as the testing set. We perform a CPA to intercept 1,000 consecutive time samples related to the masked S-box output in the first round. We use the LSB of the masked S-box output as the training label, and then use 1,000 traces of the training set to generate 1,000 new traces. 

\begin{figure}[!t]
	\centering
	\begin{minipage}[c]{0.48\textwidth}
		\centering
		\includegraphics[width=7.5cm]{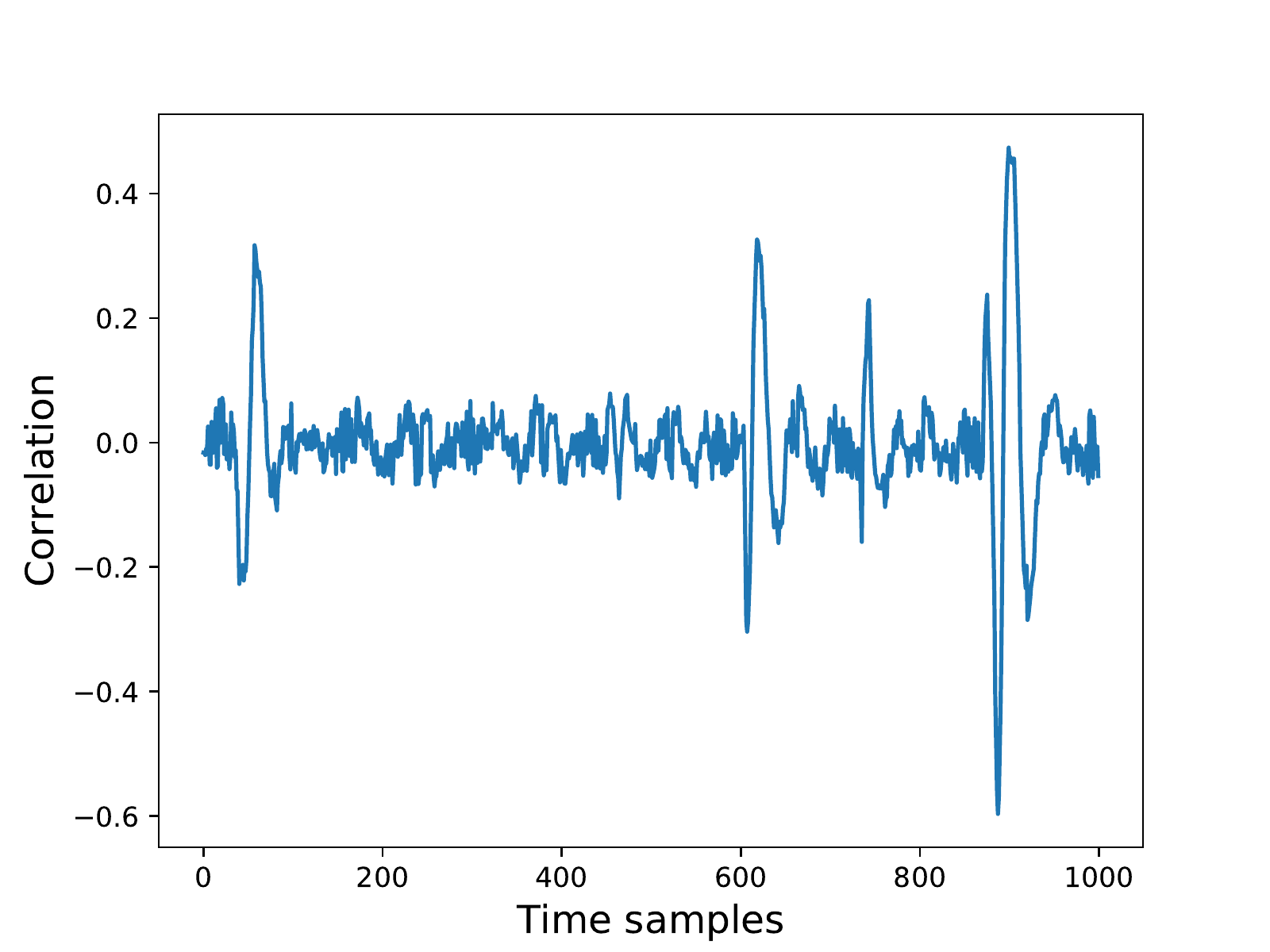}
	\end{minipage}
	\hspace{0.01\textwidth}
	\begin{minipage}[c]{0.48\textwidth}
		\centering
		\includegraphics[width=7.5cm]{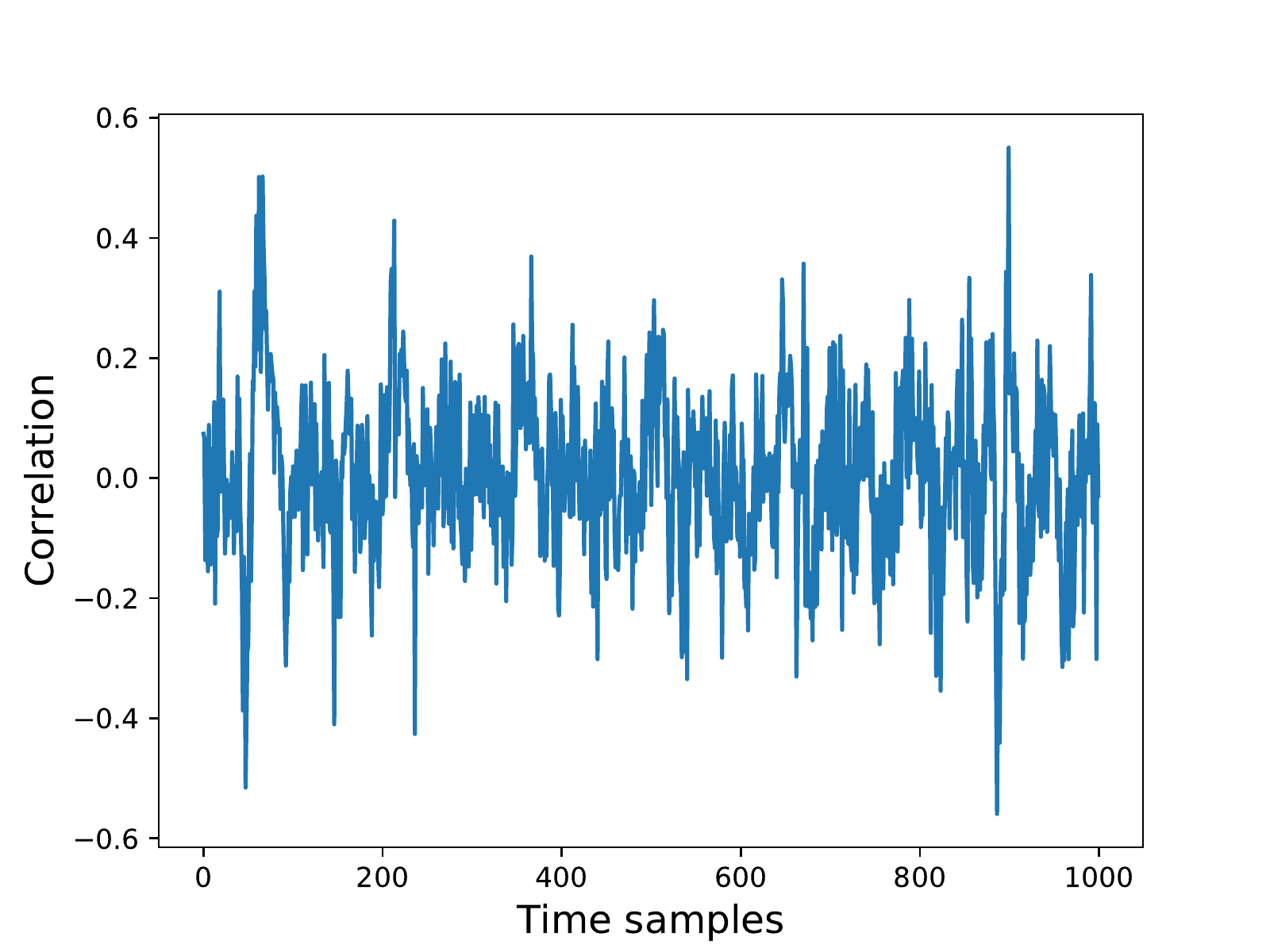}
	\end{minipage}
	\caption{Left: CPA of 1000 original traces. Right: CPA of 1000 generated traces.}
	\label{fig11}
\end{figure}

Then we use CPA to determine whether the generated traces have learned the leakages. We perform CPA with 1,000 traces of the original training set and 1,000 traces of the generated trace set respectively, the results are presented in Figure \ref{fig11}.

We can see that there are four main leakage points in the CPA results of original traces and the generated traces mainly learn the leakages of the far left and far right points, while the other two leakage points in the middle are not learned obviously. At the same time, the leakage noise at the non-leakage points is very high, thus learning the leakages of real traces is more difficult than the simulated traces. We then verify the effectiveness of the generated traces through the attack results in Table \ref{tab1}.

\begin{table}[!t]
	\footnotesize
	\caption{The attack results on DPAcontest v4.}
	\label{tab1}
	\tabcolsep 30pt 
	\begin{tabular*}{\textwidth}{cc}
		\toprule
		$\textbf{Training\ set}$ & $\textbf{Guessing\ entropy}$ \\
		\midrule
		1000 original traces & \qquad 550 \\
		1000 original traces + 1000 repeated traces & \qquad 561 \\
		1000 original traces + 1000 noisy traces & \qquad 504 \\
		1000 original traces + 1000 generated traces & \qquad \color{blue}207 \\
		2000 original traces & \qquad 172 \\
		\bottomrule
	\end{tabular*}
\end{table}

The attack results have improved significantly after using the generated traces to augment the original training set in Table \ref{tab1} and the number of traces required for GE to converge to 0 is about 300 fewer. The results confirm that the traces we generate can contribute to the model like the real traces.

\subsubsection{CW}
\label{sec:CW}

\begin{figure}[!t]
	\centering
	\begin{minipage}[c]{0.48\textwidth}
		\centering
		\includegraphics[width=7cm]{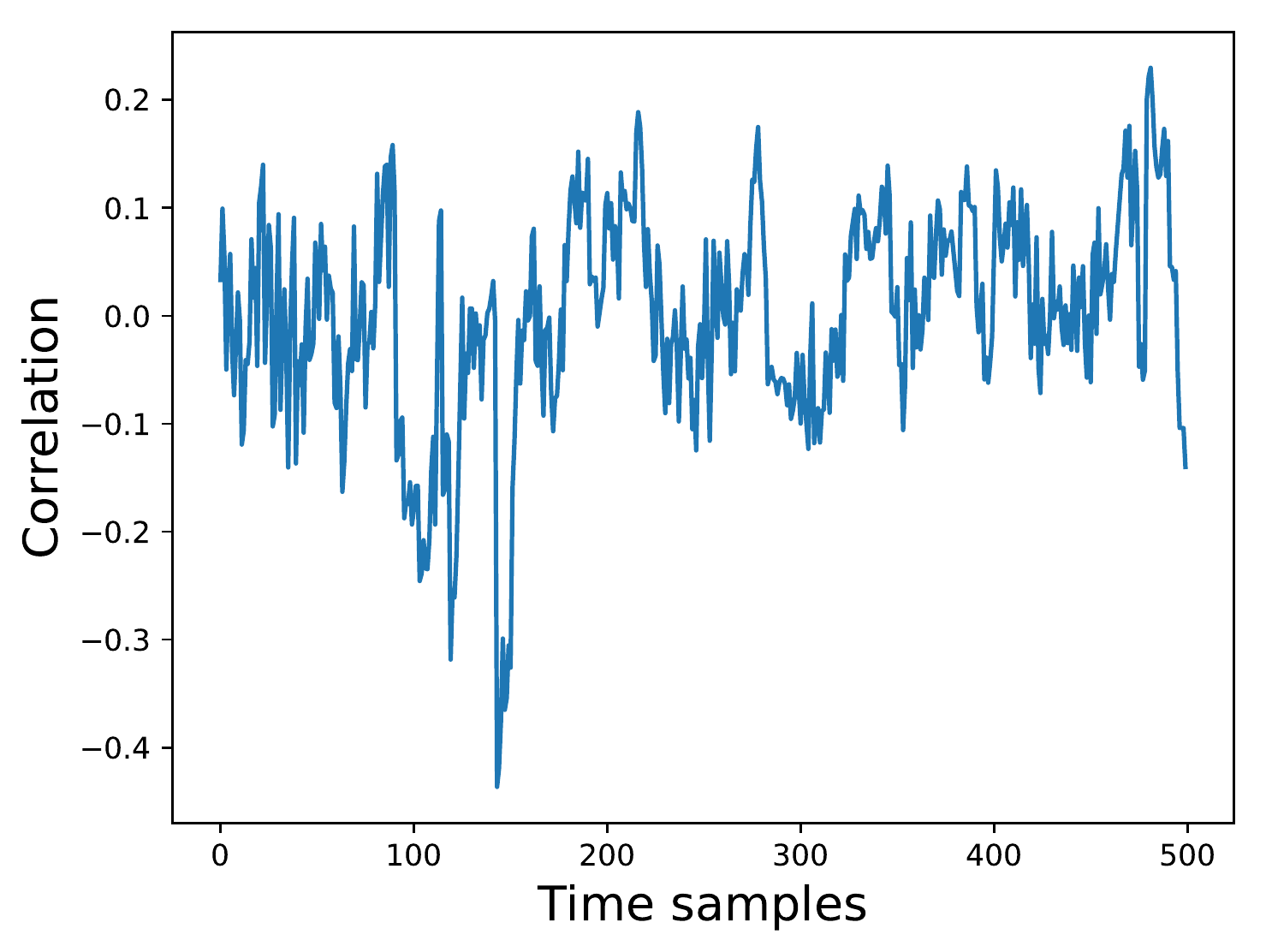}
	\end{minipage}
	\hspace{0.01\textwidth}
	\begin{minipage}[c]{0.48\textwidth}
		\centering
		\includegraphics[width=7cm]{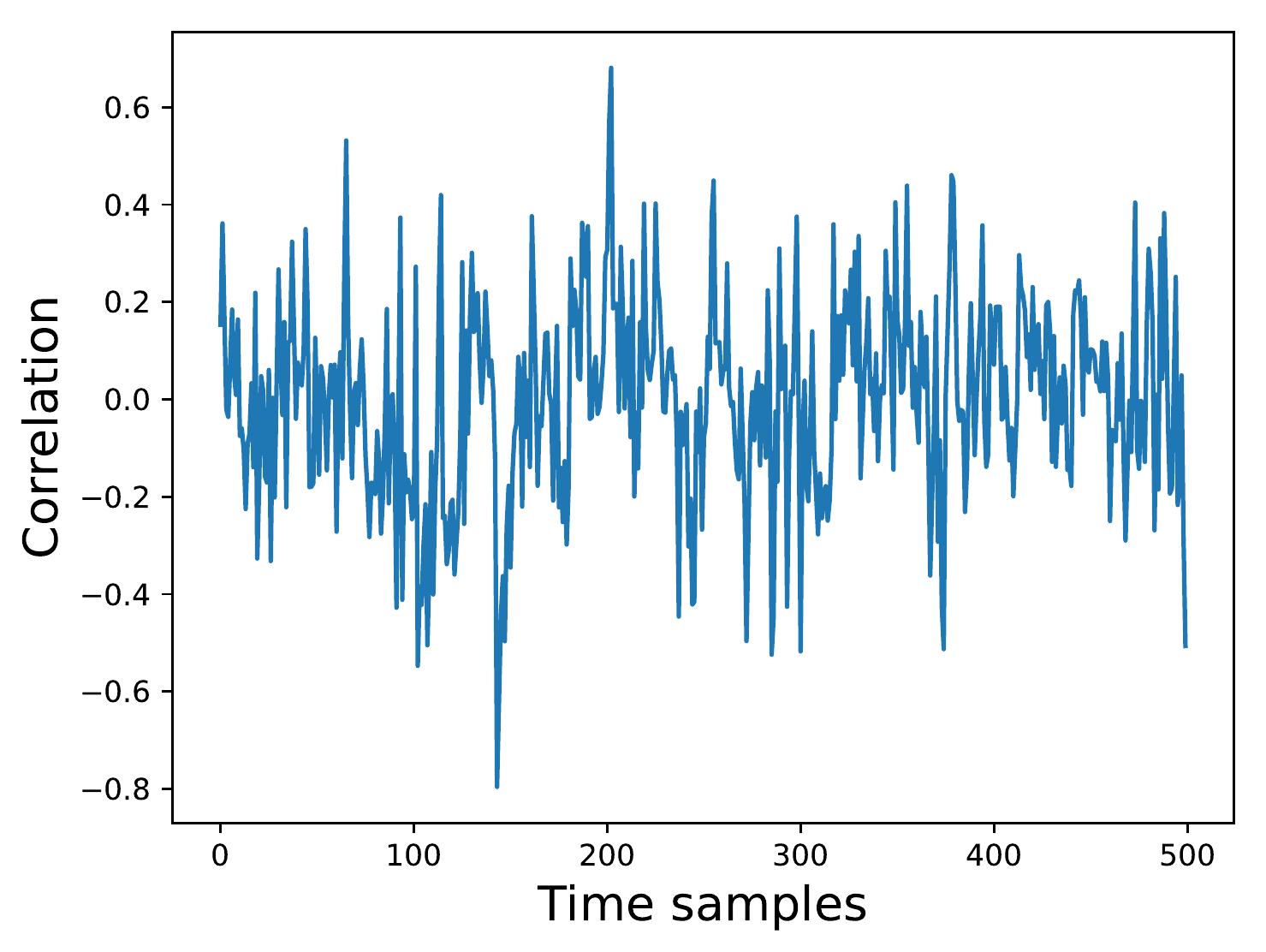}
	\end{minipage}
	\caption{Left: CPA of 100 original traces. Right: CPA of 400 generated traces.}
	\label{fig13}
\end{figure}

We use CW to collect 8000 traces to further prove the effectiveness of the generated traces. We choose the first 4000 traces as $D_1$, and the remaining 4000 traces as $D_2$. Then we randomly divide out 100 traces from $D_1$ as the training set, 2000 traces as the validation set each time, and randomly choose 2000 traces from $D_2$ as the testing set. The target of the attack is the S-box output in the first round. We intercept 500 consecutive time samples related to the first S-box output leakage points from the total 3,000 time samples in the original traces according to CPA. To simulate the profiling phase where few traces are available, we only use 100 traces as the training set. Next, we use the 100 original traces to generate 400 traces and give the comparison of CPA results in Figure \ref{fig13}.

Although the leakage noise of the generated trace is very high, the main leakage at the time samples of 100 to 200 is obvious, and we find that the overall shape of the curves in the two CPA results is similar. Then we give the attack results in Table \ref{tab2}.

\begin{table}[!t]
	\footnotesize
	\caption{The attack results on CW unprotected implementation.}
	\label{tab2}
	\tabcolsep 30pt 
	\begin{tabular*}{\textwidth}{cc}
		\toprule
		$\textbf{Training\ set}$ & $\textbf{Guessing\ entropy}$ \\
		\midrule
		100 original traces & \qquad $> 2000$ \\
		100 original traces + 400 repeated traces & \qquad $> 2000$ \\
		100 original traces + 400 noisy traces & \qquad 1859 \\
		100 original traces + 400 generated traces & \qquad \color{blue}853 \\
		500 original traces & \qquad 548 \\
		\bottomrule
	\end{tabular*}
\end{table}
From the attack results, GE can not converge to 0 by using only 100 traces as the training set when using 2000 attack traces. When we use the augmented training set to model, GE can converge to 0 by using 853 traces. Therefore, the improvement of the attack results again proves the validity of the traces generated by CGAN.

\begin{figure}[!t]
	\centering
	\includegraphics[width=.55\textwidth]{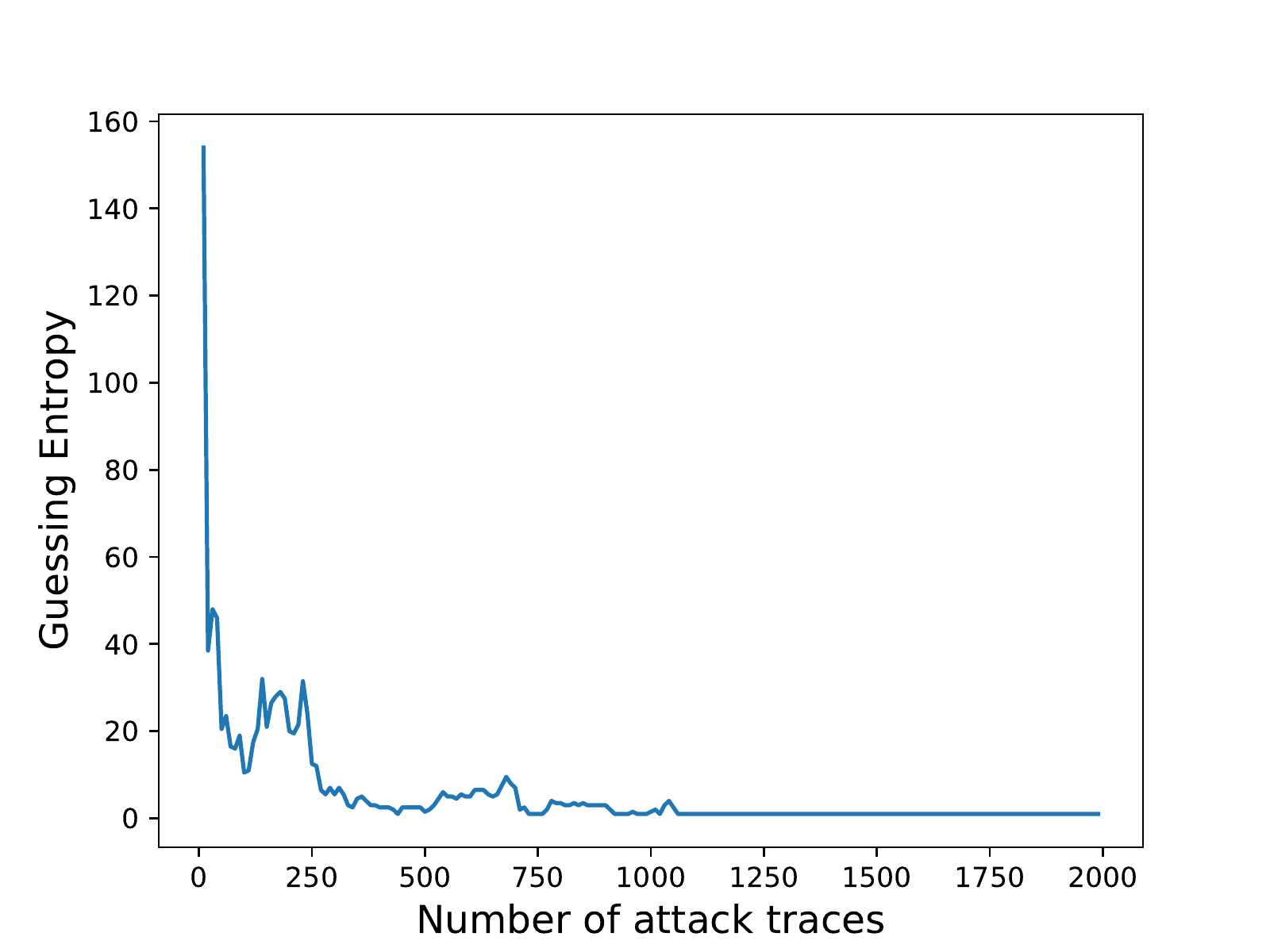}
	\caption{The attack results of model built with 400 generated traces only.}
	\label{fig15}
\end{figure}

In order to further verify that the generated traces contain useful information for modeling, we try to only use the generated traces to build a model and conduct a successful attack. Here we only use 400 generated traces from the previous 100 traces to build a model and the attack results are shown in Figure \ref{fig15}.

We can clearly see that using only 400 generated traces can attack successfully and yield a better attack performance than the original 100 traces. Therefore, the traces generated by CGAN can contribute to modeling just as real traces do. Besides, we can also evaluate the effectiveness of the generated traces by only using generated traces to build a model. 

\begin{figure}[!t]
	\centering
	\begin{minipage}[c]{0.48\textwidth}
		\centering
		\includegraphics[width=7.5cm]{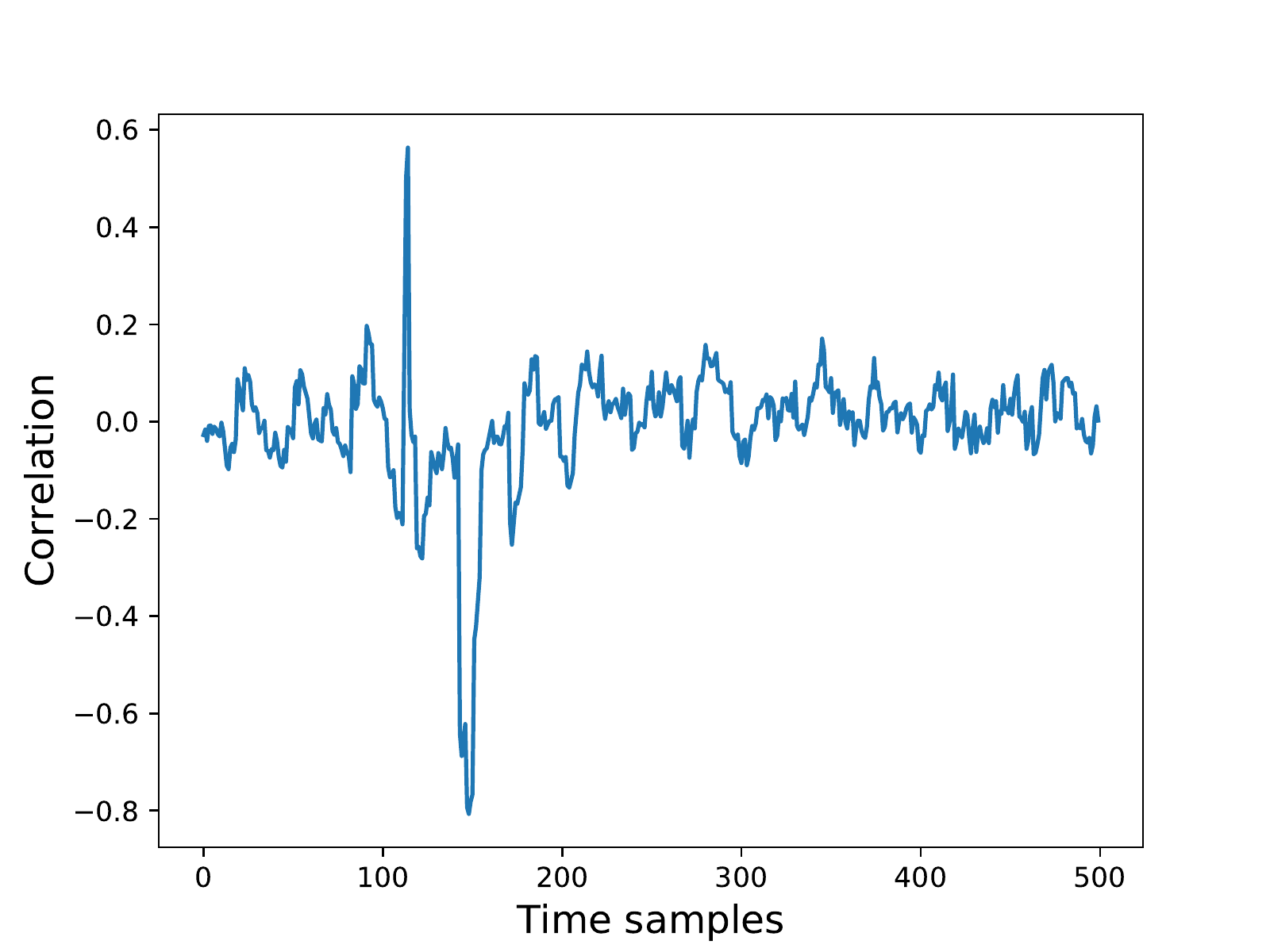}
	\end{minipage}
	\hspace{0.01\textwidth}
	\begin{minipage}[c]{0.48\textwidth}
		\centering
		\includegraphics[width=7.5cm]{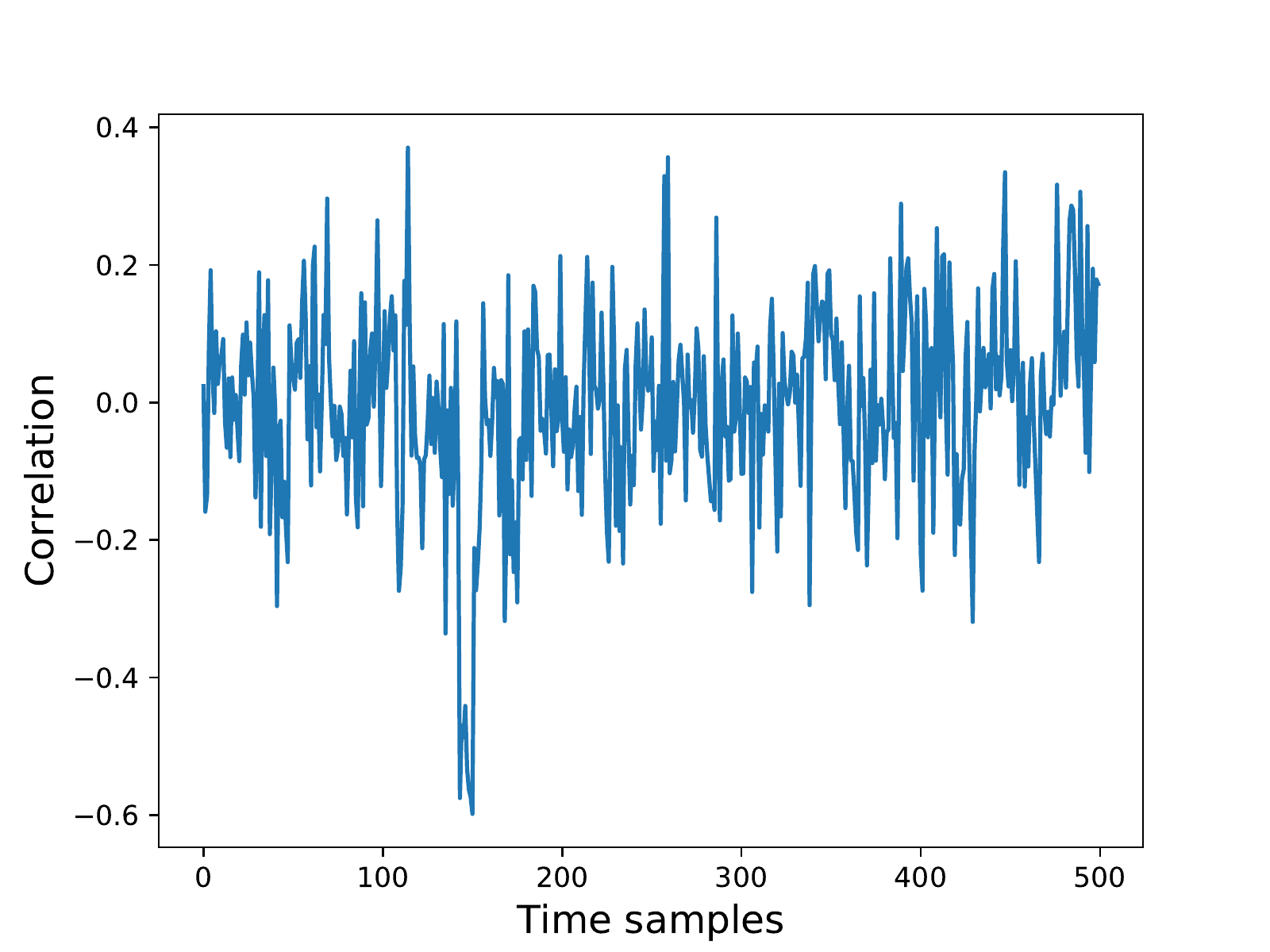}
	\end{minipage}
	\caption{Left: CPA of 400 original traces. Right: CPA of 395 generated traces.}
	\label{fig16}
\end{figure}

In \autoref{hw}, we have mentioned that using HW as a label to generate traces is difficult, especially when sufficient traces cannot be collected in the profiling phase. Here, we will test the effectiveness of generating traces using HW label in real experiments. We randomly select 400 traces from $D_1$ as the training set, 2000 traces as the validation set, and randomly choose 2000 traces from $D_2$ as the testing set. It need to be emphasized   that minority class may not appear when only 100 traces are used to generate. Then we use 400 traces in the training set to generate 395 traces which still obey the binomial distribution. First, we perform CPA on the training set and the generated traces respectively. The CPA results are presented in Figure \ref{fig16}.

\begin{figure}[!t]
	\centering
	\includegraphics[width=.55\textwidth]{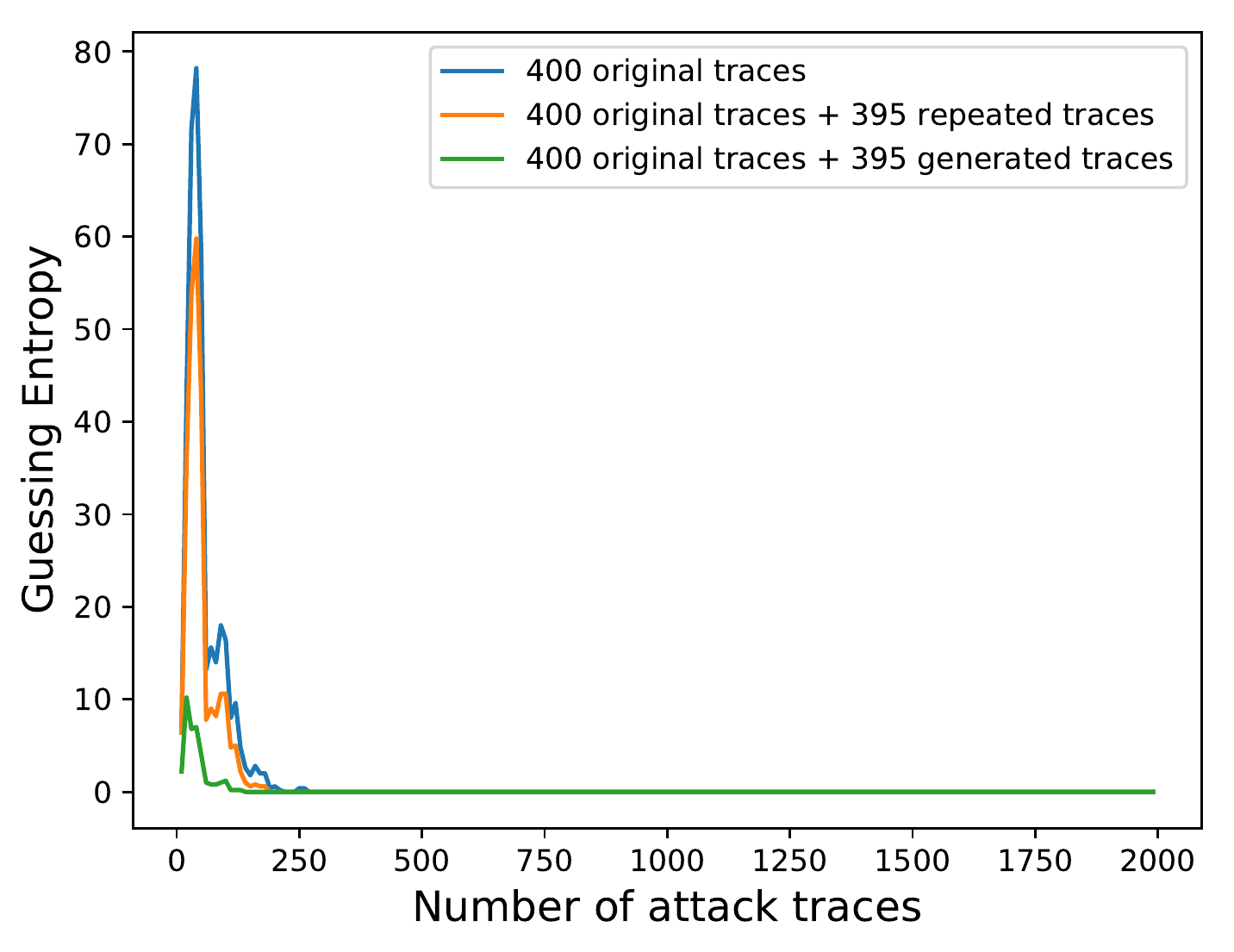}
	\caption{The attack results of models built with \{400 original traces, 400 original traces + 395 repeated traces, 400 original traces + 395 generated traces\} when using HW as label.}
	\label{fig17}
\end{figure}

The correlation of the generated traces at the main leakage points has decreased, which was also mentioned in previous simulation experiments. Next, we respectively use the original training set and the training set augmented with the generated traces to model. The attack results are presented in Figure \ref{fig17}.

It can be seen that in real experiments, it is feasible to use the HW label to generate traces, but the improvement of the attack performance is not obvious. This also reflects that it is not appropriate to use HW as a label to generate traces when profiling traces are insufficient.

\subsection{First-Order Masked AES Implementation}
\label{sec:mask}

In this section, we try to expand the application scope of using CGAN to generate traces and test whether it can still improve the performance of profiling attacks on first-order masked AES. Here we use dataset ASCAD.h5 in the public database ASCAD. One purpose of the paper \cite{DBLP:journals/iacr/ProuffSBCD18} is to investigate whether a DL model can break masking scheme protection. Therefore, it does not calculate the output of the masked S-box, but directly takes the output of the third S-box as the label. In fact, there is no first-order leakage. Hence, when we generate traces using the label directly related to the S-box output, we can not see obvious peaks in the CPA results of generated traces. Although it is not possible to judge whether the generated traces have learned leakages through CPA, we can use the attack results to verify the effectiveness of the generated traces.

We let the training set of 50000 traces be $D_1$, and the testing set of 10000 traces be $D_2$. In order to simulate a scenario where sufficient traces cannot be collected in practical profiling phase, we randomly select 2000 traces from $D_1$ as the training set, 2000 traces as the validation set, and randomly select 2000 traces from $D_2$ as the testing set. The label is $LSB(Sbox(p\left[3\right]\oplus r\left[3\right]))$. Then we use the training set to generate 2000 traces.

\begin{figure}[!t]
	\centering
	\begin{minipage}[c]{0.48\textwidth}
		\centering
		\includegraphics[width=7cm]{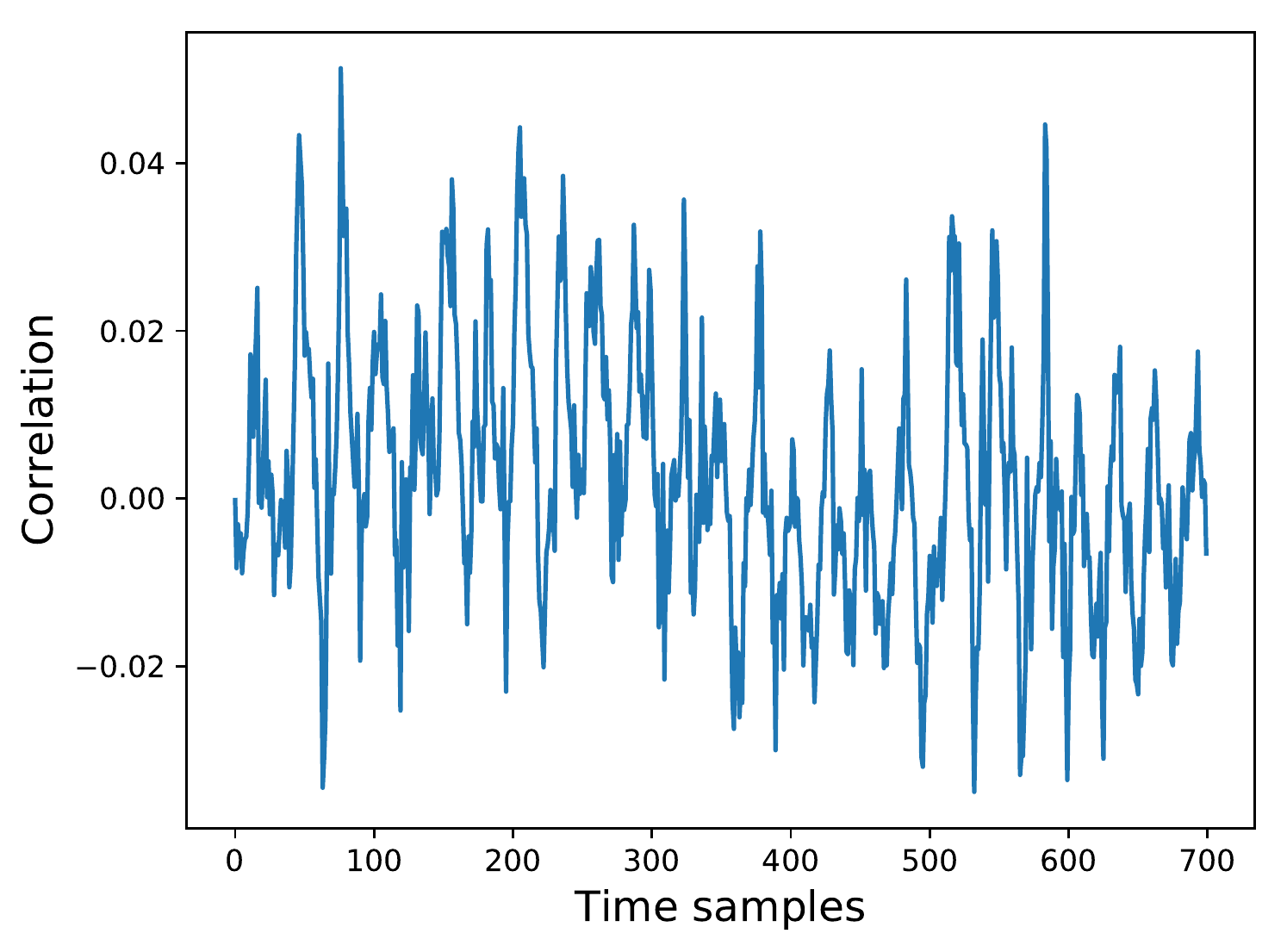}
	\end{minipage}
	\hspace{0.01\textwidth}
	\begin{minipage}[c]{0.48\textwidth}
		\centering
		\includegraphics[width=7cm]{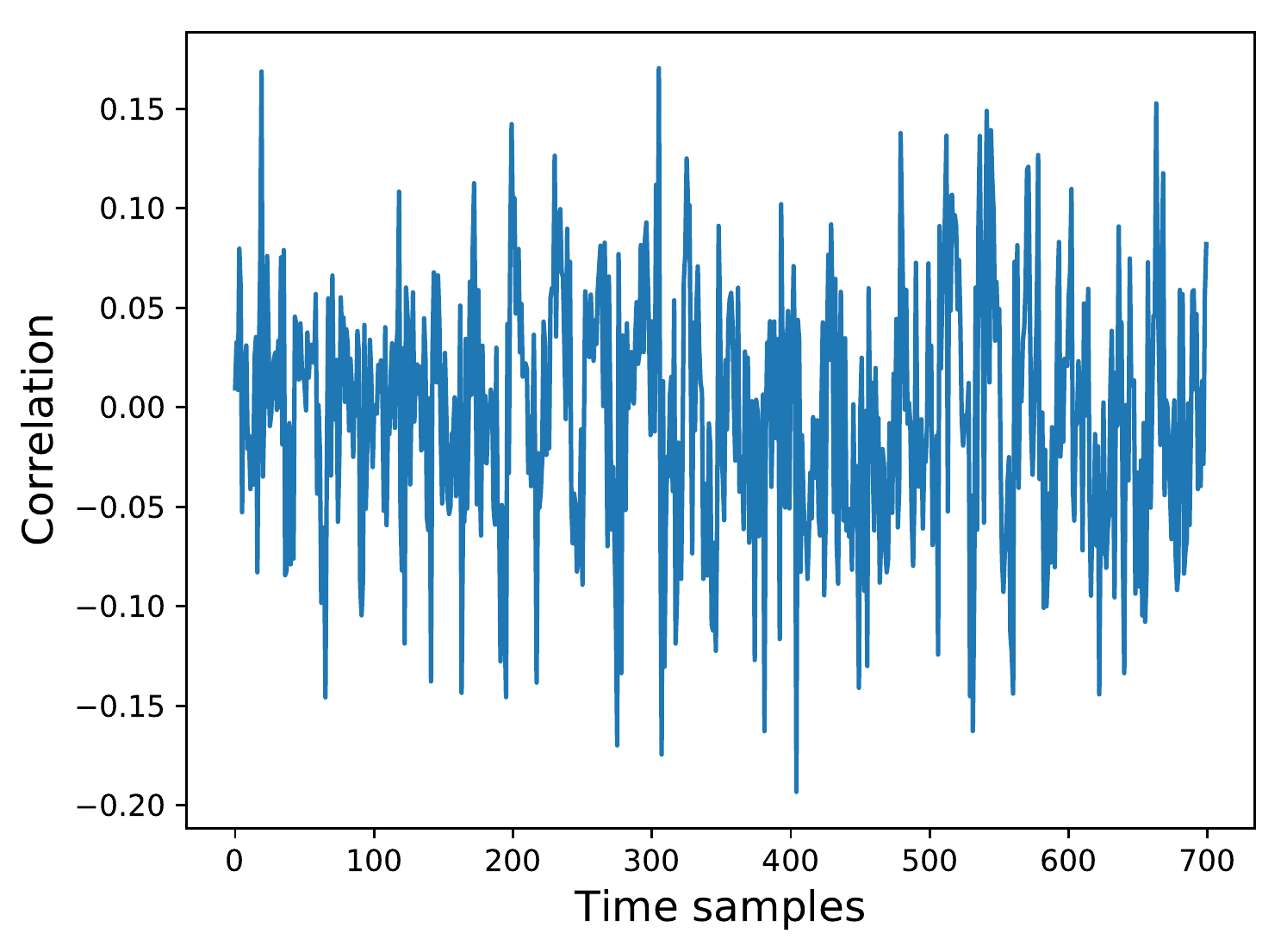}
	\end{minipage}
	\caption{Left: CPA of 100 original traces. Right: CPA of 395 generated traces.}
	\label{fig18}
\end{figure}

First, we still compare the CPA results of the original training set and generated traces. The results are presented in Figure \ref{fig18}. As mentioned above, the S-box output of the first-order masked AES has no first-order leakage, and no leakages can be seen in the CPA results of the original training set or the generated traces. 
Then we judge the effectiveness of the generated traces by whether it can improve the attack performance of the model. The comparison of attack results is shown in Table \ref{tab3}.

\begin{table}[!t]
	\footnotesize
	\caption{The attack results on ASCAD.h5.}
	\label{tab3}
	\tabcolsep 30pt 
	\begin{tabular*}{\textwidth}{cc}
		\toprule
		$\textbf{Training\ set}$ & $\textbf{Guessing\ entropy}$ \\
		\midrule
		2000 original traces & 878 \\
		2000 original traces + 2000 repeated traces & 852 \\
		2000 original traces + 2000 noisy traces & 753 \\
		2000 original traces + 2000 generated traces & \color{blue}182 \\
		4000 original traces & 91 \\
		\bottomrule
	\end{tabular*}
\end{table}
Compared with the original training set, the training set expanded with the generated traces can build a better model, and the traces required for GE to converge to 0 is reduced by almost four times. In order to further prove that CGAN can learn higher-order leakage, we conduct a more detailed trace generation using known masks. As shown in the following table, we divide the generated traces into four categories in Table \ref{tab4}.

\begin{table}[!t]
	\footnotesize
	\caption{Based on the known masks, we construct four labels for the training set of CGAN.}
	\label{tab4}
	\tabcolsep 22pt 
	\begin{tabular*}{\textwidth}{cccc}
		\toprule
		$LSB(mask)$ & $LSB(masked)$ & $Gan\_label$ & $LSB(Sbox\_output)$\\
		\midrule
		0 & 0 & 0 & 0 \\
		1 & 0 & 1 & 1 \\
		0 & 1 & 2 & 1 \\
		1 & 1 & 3 & 0 \\
		\bottomrule
	\end{tabular*}
\end{table}
Four labels in the Table \ref{tab4} have the following relationships: 
\begin{equation}
Gan\_label=2\cdot LSB(masked)+LSB(mask).
\label{eq6}
\end{equation}
\begin{equation}
LSB(Sbox\_output)=LSB(mask)\oplus LSB(masked).
\label{eq7}
\end{equation}

\begin{figure}[!t]
	\centering
	\begin{minipage}[c]{0.48\textwidth}
		\centering
		\includegraphics[width=7.4cm]{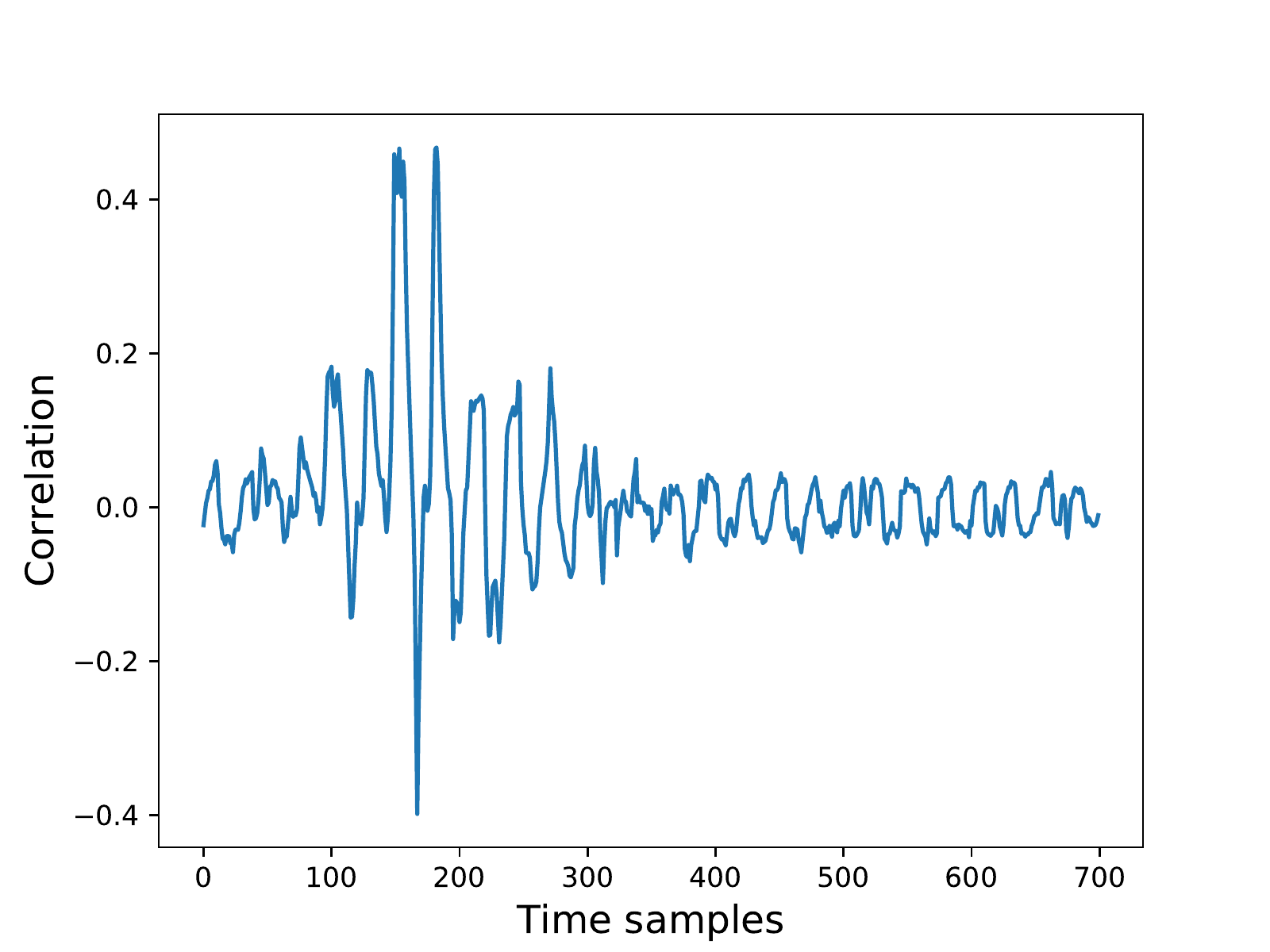}
	\end{minipage}
	\begin{minipage}[c]{0.48\textwidth}
		\centering
		\includegraphics[width=7.4cm]{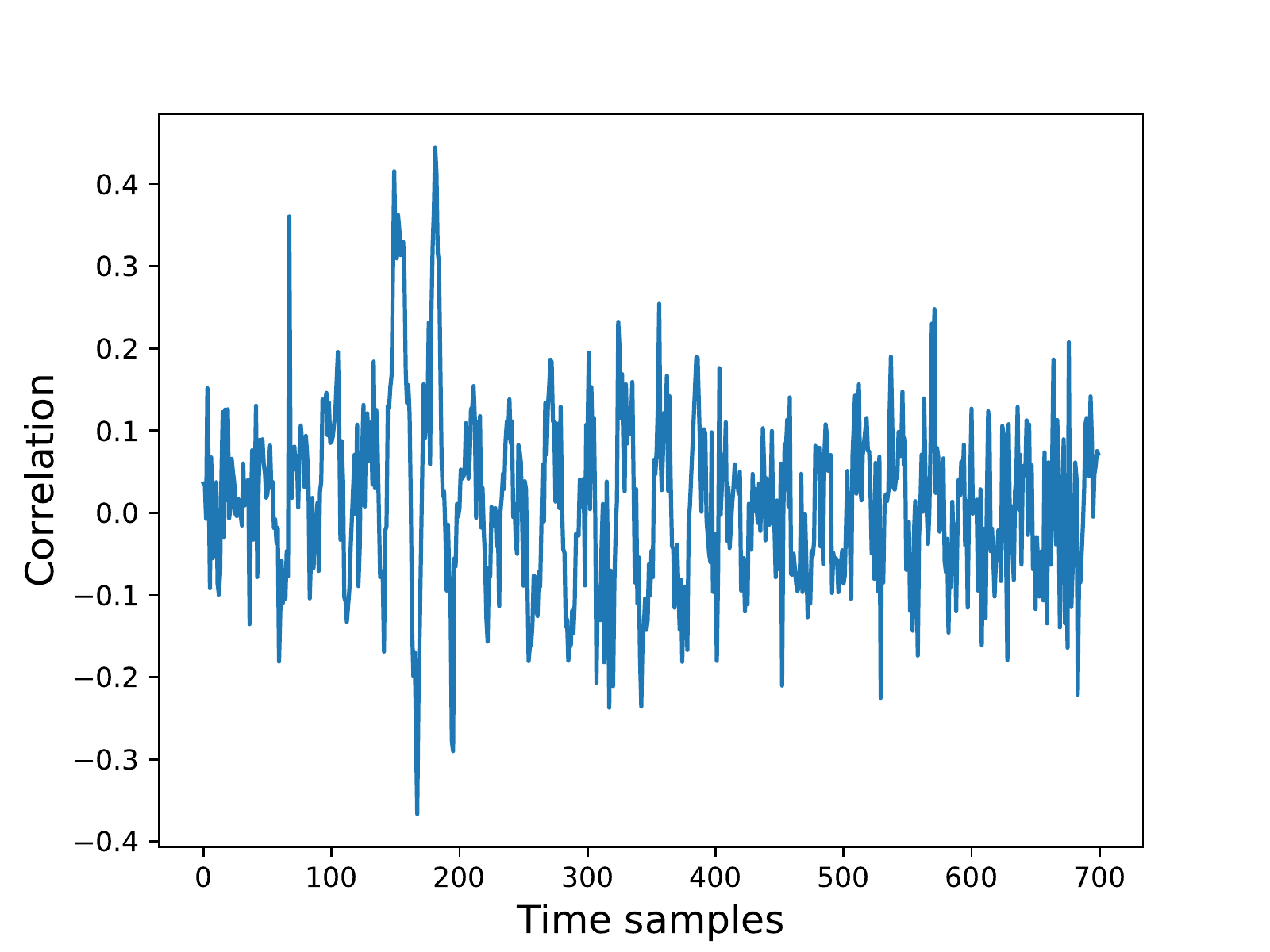}
	\end{minipage}
	
	\begin{minipage}[c]{0.48\textwidth}
		\centering
		\includegraphics[width=7.4cm]{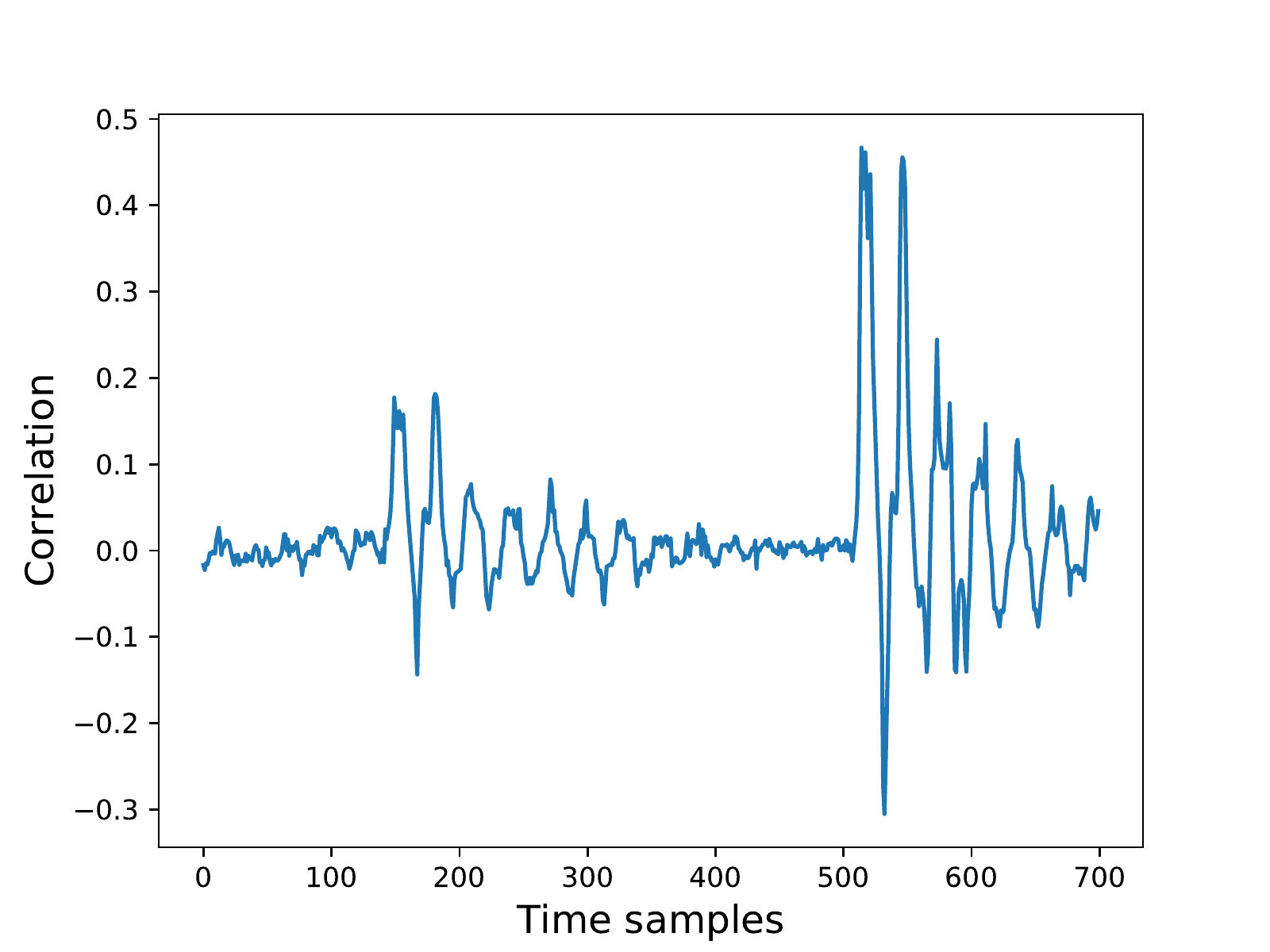}
	\end{minipage}
	\begin{minipage}[c]{0.48\textwidth}
		\centering
		\includegraphics[width=7.44cm]{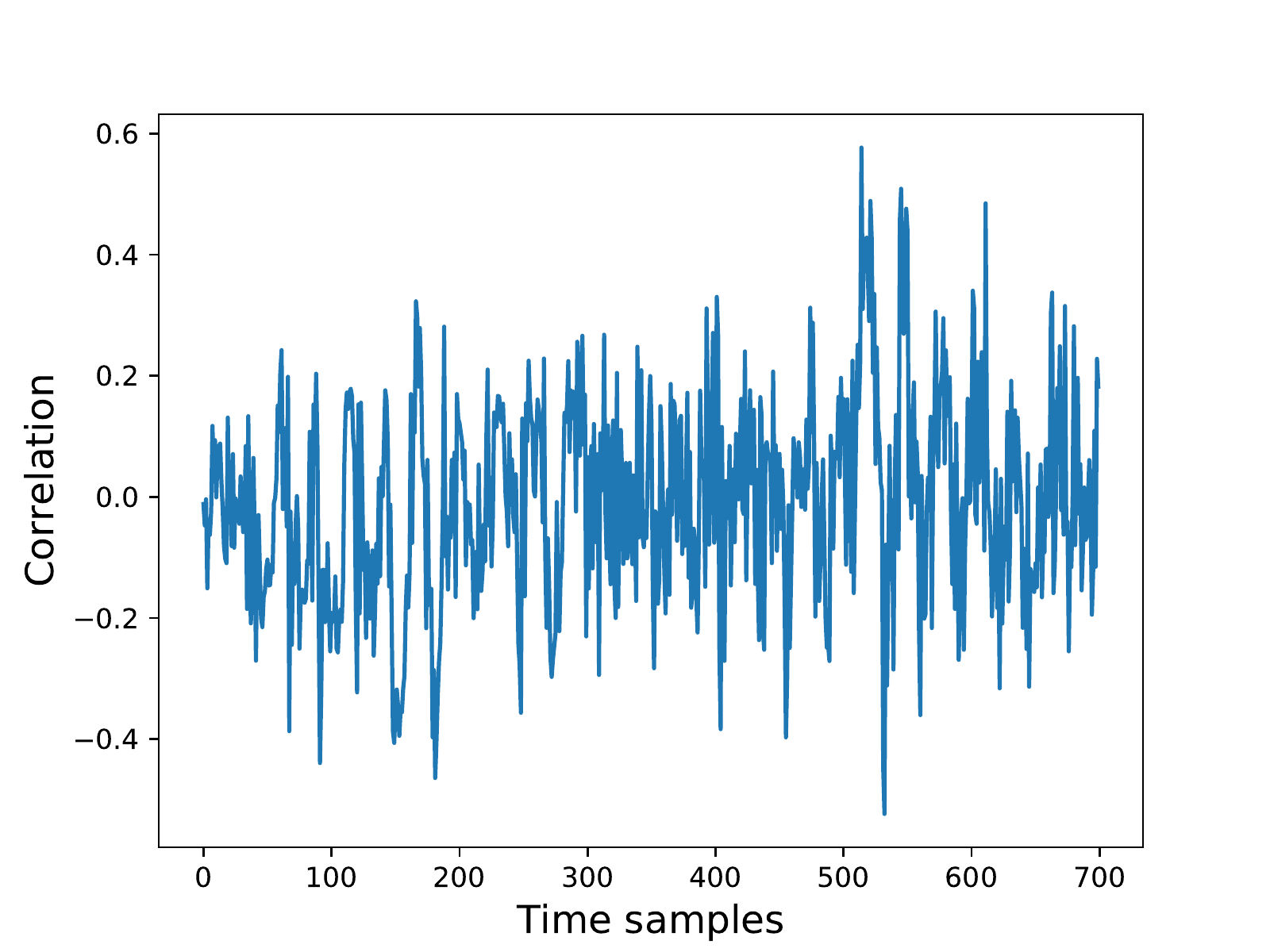}
	\end{minipage}
	\caption{Top-left: mask CPA of original traces. Top-right: mask CPA of generated traces. Bottom-left: masked CPA of original traces. Bottom-right: masked CPA of generated traces.}
	\label{fig20}
\end{figure}

Based on Eq. (\ref{eq6}), we can get $LSB(mask)$ and $LSB(masked)$ from $Gan\_label$\footnote{$Gan\_label$ refers to the label used during CGAN training.}, to perform a CPA on the generated traces targeting the mask and the masked S-box output. On the other hand, we can calculate the $LSB(Sbox\_output)$ from $LSB(mask)$ and $LSB(masked)$ through Eq. (\ref{eq7}), as the label of generated traces when expanding the original training set. First, we randomly divide out 2000 traces from $D_1$ as a training set. Then we generate 2,000 traces using the original training set, with 500 traces in each class. Finally, the results of CPA targeting mask and masked S-box output which is performed on the original training set and generated traces respectively are displayed in Figure \ref{fig20}:

From above CPA results, the generated traces can learn major leakage related to both mask and masked S-box output, which indicates that CGAN can learn high-order leakages. At the same time, this experiment is also a supplement to previous trace generation only using LSB of the S-box output. It confirms that the generated traces in the previous experiment have learned effective leakage. Therefore, even if the mask cannot be obtained in a real situation, we can directly use the label related to the output of the S-box to effectively generate traces.

\subsection{Desynchronized Traces}
\label{sec:desync}

In this section, we continue to study the application of using CGAN to generate traces on the desynchronized traces. We first perform a simulation experiment to verify whether CGAN can generate effective desynchronized traces. The configuration of the simulation experiment is the same as that in section \ref{sec:sim}, except that we add a random jitter $s \sim(-2,2)$ to the original leakage point. The location of the leakage point has changed to $t_{leakage}=25+s$, thus we obtain a desynchronized dataset.

\begin{figure}[!t]
	\centering
	\begin{minipage}[c]{0.48\textwidth}
		\centering
		\includegraphics[width=7.5cm]{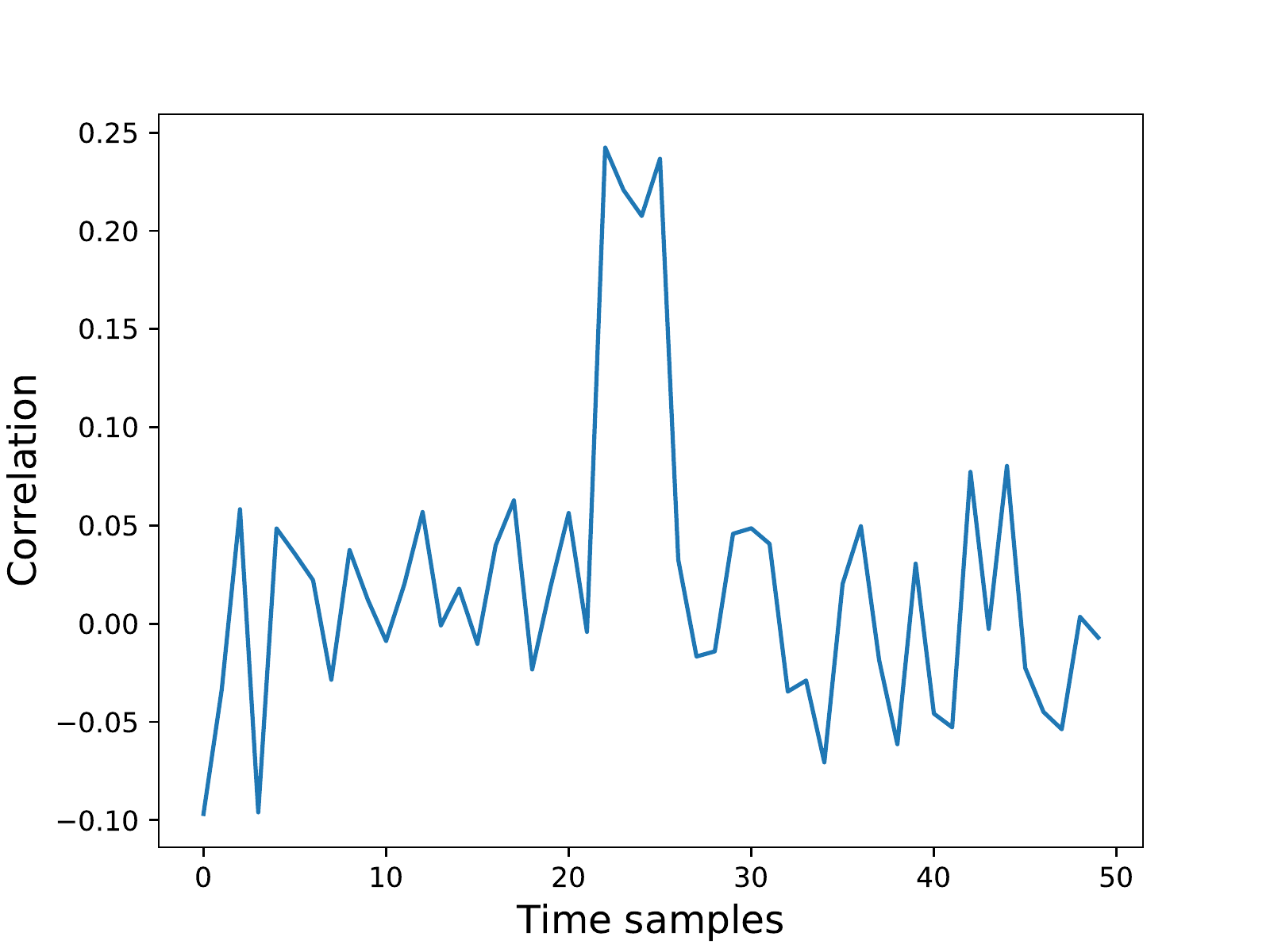}
	\end{minipage}
	\hspace{0.01\textwidth}
	\begin{minipage}[c]{0.48\textwidth}
		\centering
		\includegraphics[width=7.5cm]{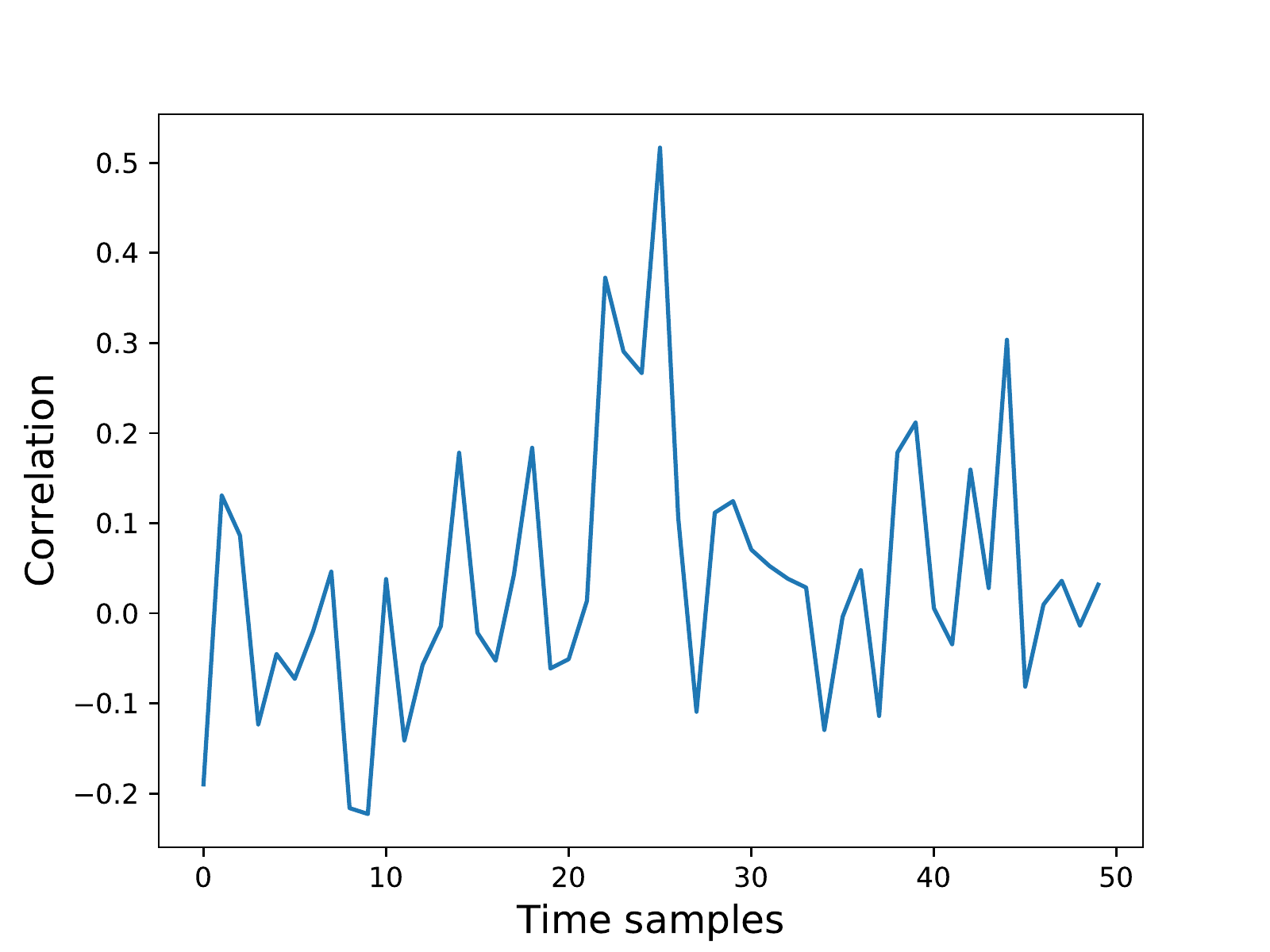}
	\end{minipage}
	\caption{Left: CPA of 500 original traces. Right: CPA of 400 generated traces.}
	\label{fig21}
\end{figure}

In each repeated experiment, the 500 traces of the original training set are still used to generate 400 traces. The CPA results of the generated traces and the original traces are shown in Figure \ref{fig21}.

\begin{figure}[!t]
	\centering
	\includegraphics[width=.55\textwidth]{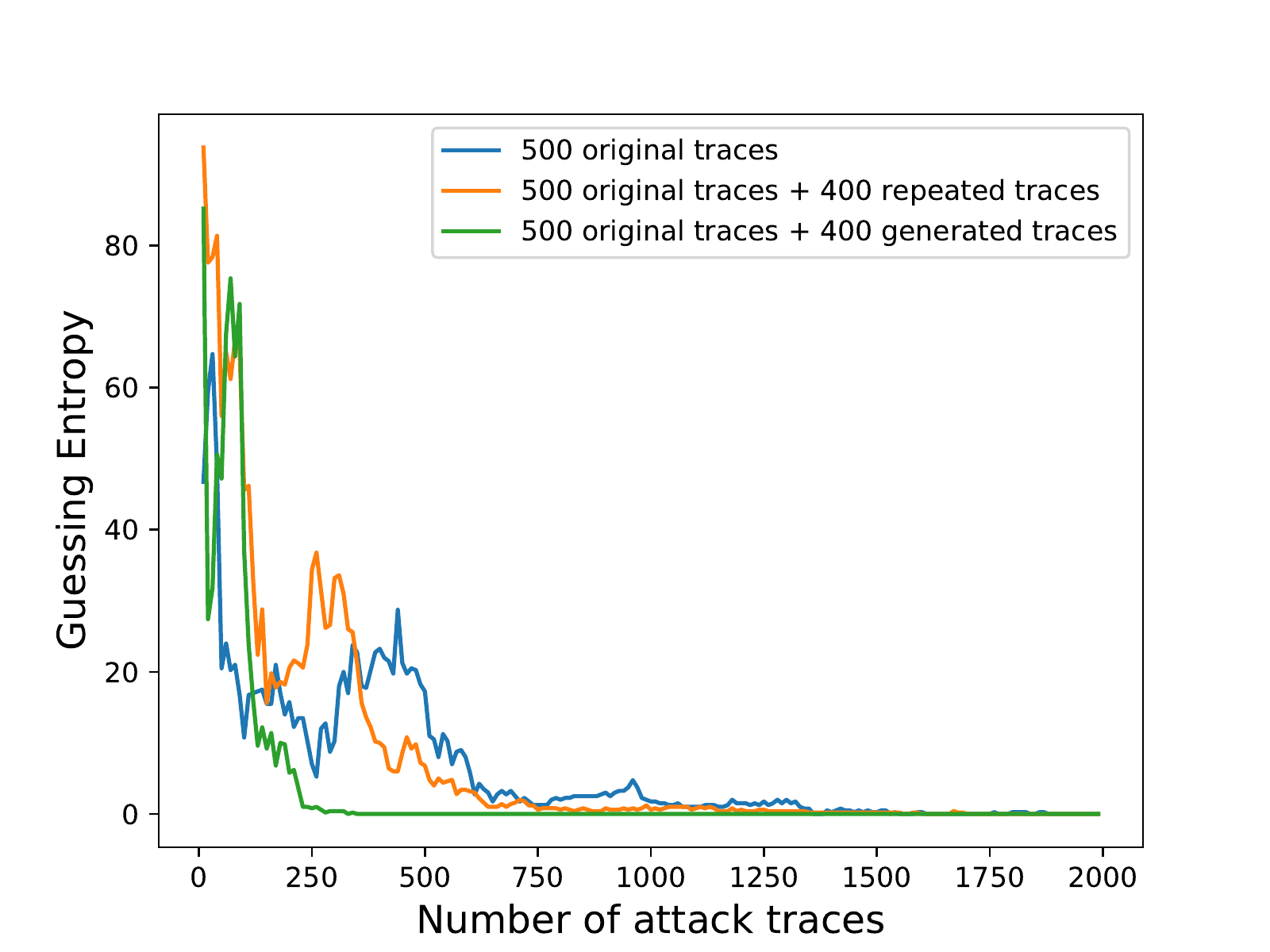}
	\caption{The attack results of models built with \{500 original traces, 500 original traces + 400 repeated traces, 500 original traces + 400 generated traces\} on the simulated desynchronization traces.}
	\label{fig22}
\end{figure}

As a result of desynchronization, the correlation peak at the leakage point of the original training set not only drops a lot but also becomes wider. While in the CPA results of generated traces, we find that the correlation at the leakage point is twice as high, but the leakage noise is also more obvious, which indicates that it is more difficult to generate desynchronized traces. Next, we give a comparison of the attack results in Figure \ref{fig22}.

The attack performance of the trained model based on desynchronized traces is much worse than the synchronized traces in section \ref{sec:sim}. Nevertheless, using the generated traces to augment the original trace set can still effectively improve the attack performance.

Next, we will continue to verify the effectiveness of CGAN on desynchronized traces by using AES\_RD dataset which is collected in the AES implementation with random delay countermeasure \cite{DBLP:conf/ches/CoronK09}. The original AES\_RD dataset contains 50,000 traces. We choose the first 30,000 traces as $D_1$, the last 20,000 traces as $D_2$. Then we randomly divide out 10,000 traces from $D_1$  as the training set and 10,000 traces as the validation set.
We randomly divide out 10,000 traces from $D_2$ as the testing set. The label we use is also LSB. Finally, we use the training set to generate 10,000 traces. The attack results are presented in Table \ref{tab5}.

\begin{table}[!t]
	\footnotesize
	\caption{The attack results on AES\_RD.}
	\label{tab5}
	\tabcolsep 30pt 
	\begin{tabular*}{\textwidth}{cc}
		\toprule
		$\textbf{Training\ set}$ & $\textbf{Guessing\ entropy}$ \\
		\midrule
		10000 original traces & \qquad 4125 \\
		10000 original traces + 10000 repeated traces & \qquad 3998 \\
		10000 original traces + 10000 noisy traces & \qquad 3365 \\
		10000 original traces + 10000 generated traces & \qquad \color{blue}1369 \\
		20000 original traces & \qquad 890 \\
		\bottomrule
	\end{tabular*}
\end{table}

The number of traces required for a successful attack has been reduced by about 2700 when we use the original training set augmented with generated traces to build a model. And it also shows that CGAN can generate effective traces even when the traces are not synchronized.

\subsection{Application to Other Modeling Algorithms}

In previous experiments, the model we have always used is MLP. In this section, we will verify the universality of our method to different modeling algorithms. We test three models, SVM, RF, and CNN. The first two are machine learning models, and the last one is a DL model which is often applied to profiling attacks in recent years.

Here we will test the datasets used in sections \ref{sec:CW} and \ref{sec:mask} and the experimental configuration is the same as before, except that the modeling algorithm used is different. The specific parameters used in the model are given in Appendix \ref{appendix_classifier}. Then, under three other modeling algorithms, we compare the attack results before and after using the generated traces to expand the original training set.

\begin{table}[!t]
	\footnotesize
	\caption{The attack results of model SVM, RF, CNN on dataset CW, ASCAD.}
	\label{tab6}
	\tabcolsep 12pt 
	\begin{tabular*}{\textwidth}{ccccc}
		\toprule
		$\textbf{Dataset}$ & $\textbf{Training\ set}$ & $\textbf{SVM}$ & $\textbf{RF}$ & $\textbf{CNN}$ \\
		\midrule
		
		\multirow{5}{*}{CW} & 100 original traces & $>2000$ & 250 & $>2000$ \\
		& 100 original traces + 400 repeated traces & 1305 & 259 & $>2000$ \\
		& 100 original traces + 400 noisy traces & 1591 & 204 & 1760 \\
		& 100 original traces + 400 generated traces & \color{blue}316 & \color{blue} 76 & \color{blue} 171\\
		& 500 original traces & 194 & 53 & 86 \\
		\midrule
		\multirow{5}{*}{ASCAD} & 2000 original traces & 1502 & 625 & 850 \\
		& 2000 original traces + 2000 repeated traces & 1493 & 874 & 702 \\
		& 2000 original traces + 2000 noisy traces & 941 & 533 & 597 \\
		& 2000 original traces + 2000 generated traces & \color{blue}313 & \color{blue} 380 & \color{blue} 294\\
		& 4000 original traces & 244 & 257 & 232 \\
		\bottomrule
	\end{tabular*}
\end{table}

From the attack results in Table \ref{tab6}, we can see that even if different modeling algorithms are used, in scenarios where the profiling traces are insufficient, the traces generated by CGAN can improve the attack performance. This also confirms the universality of our method to different modeling algorithms.

\section{Discussions}
\label{diss}
\begin{enumerate}[leftmargin=\parindent, itemsep=0.1pt]
	\item  \textbf{Compared with the previous profiling attack work, our attack effect improvement seems to be smaller.}
	We want to emphasize that our work is in a scene with insufficient profiling traces, so it is very difficult to improve the attack effect. At the same time, when we compare some other data augmentation techniques, such as adding Gaussian noise, the improvement effect is very weak, which also shows that the generation of the adversarial network is powerful and can produce traces that bring additional useful information. Of course, we are manually simulating the lack of traces scenarios to confirm the effectiveness of CGAN and provide a method to solve the problem of how to better improve profiling attacks in this scenario.
	\item \textbf{In the scenario of insufficient profiling traces, the attack effect of the constructed model is not good, does it mean that the model is not optimal?}
	One thing to note is that the optimal model we use here is relatively in a scene with sufficient trace, just like the public model we used in this paper. From the perspective of information theory, in a scenario where the profiling traces are lacking, the current model cannot extract more information, and a better model may exist. Our work is not to directly optimize this model, but to generate new traces so that existing profiling methods can get better profiling.
\end{enumerate}

\section{Conclusion}
\label{sec:conc}

In this paper, we first introduce CGAN(GAN) in the context of side-channel attacks. When sufficient traces cannot be collected during the profiling phase, CGAN can generate traces for data augmentation, which effectively improves the performance of profiling attacks. In the process of using CGAN to generate traces, we summarize some methods and techniques: By CPA or DPA, we can determine in advance whether the generated traces have learned effective leakages. It is easy to use the single-bit label to generate effective traces. We must add a suitable amount of generated traces when augmenting the original profiling set. 
We find that CGAN has the ability to learn not only first-order leakage, but also high-order leakage, and it can also effectively generate desynchronized traces. 
Besides, we also test different modeling algorithms, including SVM, RF, MLP, and CNN, which demonstrates the universality of our method to different modeling algorithms.
In future work, we plan to generate traces collected from other types of cryptographic implementations by CGAN and improve the quality of generated traces.

\bibliographystyle{alpha}
\bibliography{reference}

\appendix

	\section{}
	\subsection{CGAN}
	\label{appendix_cgan}
	
	\begin{table}[h]
		\footnotesize
		\tabcolsep 10pt 
		\begin{tabular*}{\textwidth}{ccc}
			\toprule
			& \textbf{Generator} & \textbf{Discriminator}\\
			\midrule
			\textbf{Input layer} & (noise with $laten\_dim=100$,embedding(label)) & (traces, embedding(label)) \\
			\midrule
			\textbf{Hidden layer} & (Dense,LeakyReLU,BN) * N & (Dense, LeakyReLU) * N\\
			\midrule
			\textbf{Output layer} & Dense with tanh & Dense with sigmoid\\
			\bottomrule
		\end{tabular*}
	\end{table}
	
	\subsection{DPA Analysis}
	\label{dpa}
\begin{figure}[h]
			\centering
			\begin{minipage}[c]{0.48\textwidth}
				\centering
				\includegraphics[width=7cm]{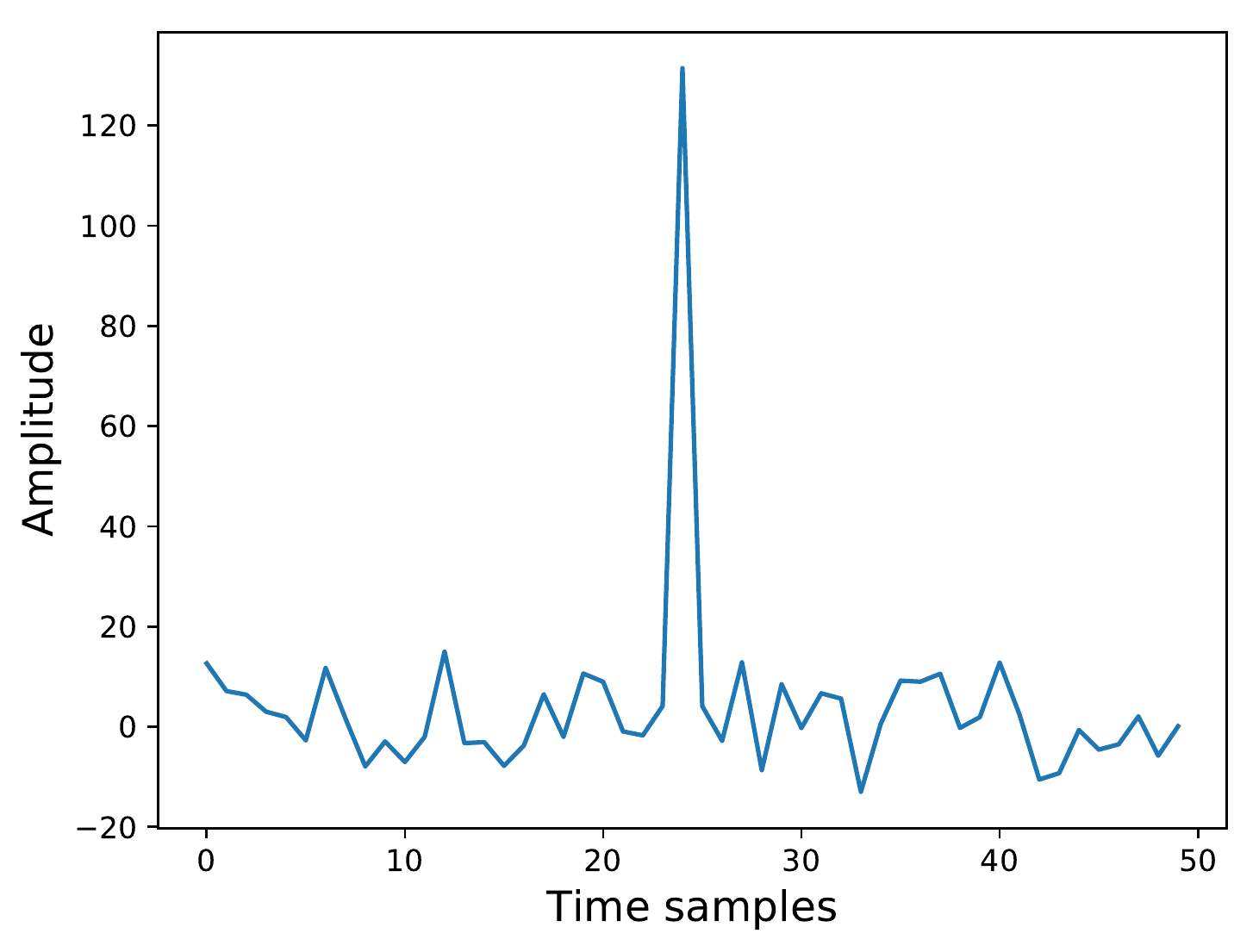}
			\end{minipage}
			\hspace{0.01\textwidth}
			\begin{minipage}[c]{0.48\textwidth}
				\centering
				\includegraphics[width=7cm]{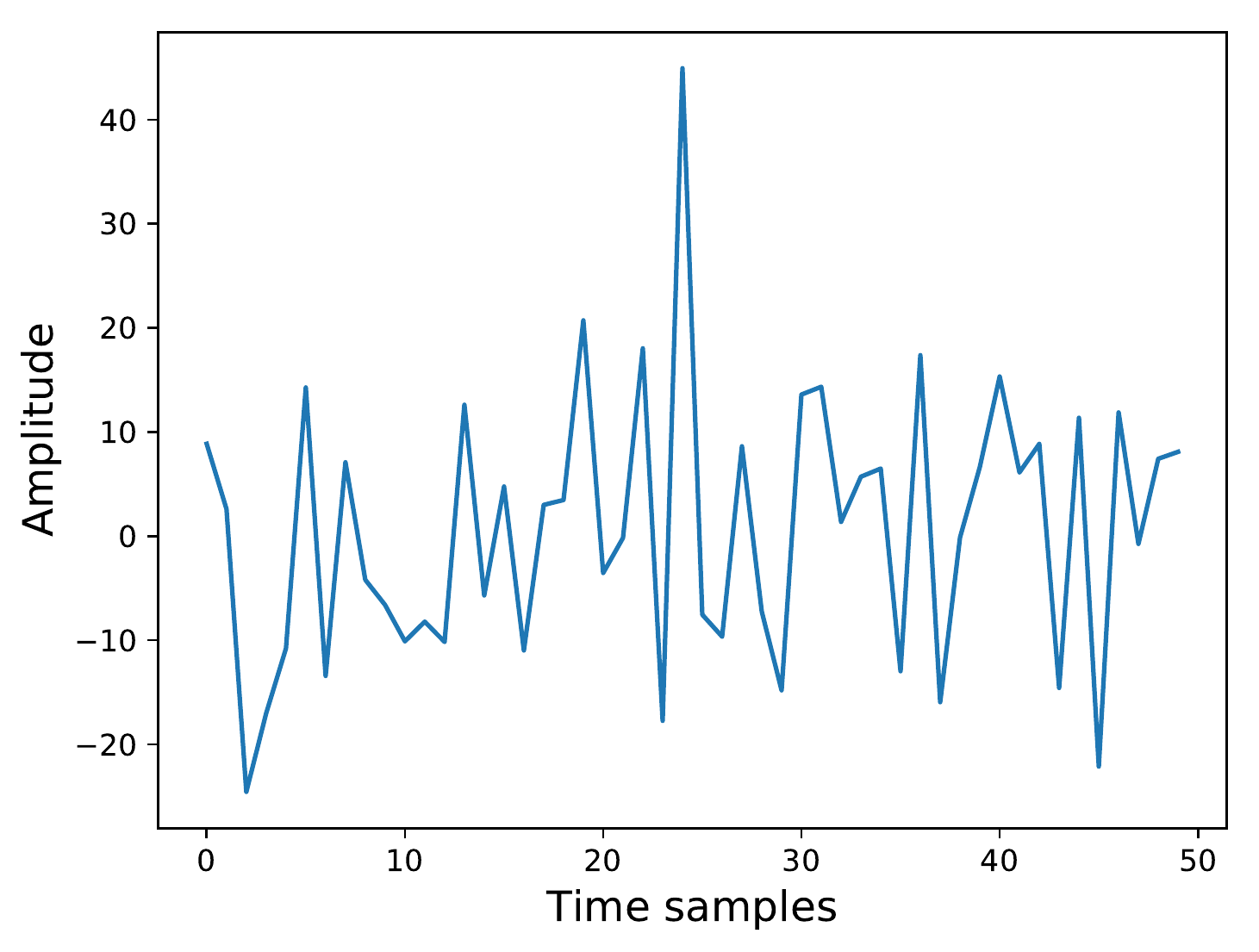}
			\end{minipage}
			\caption{Left: DPA of 500 original traces. Right: DPA of 400 generated traces.}
			\label{sim_dpa}
		\end{figure}

\begin{figure}[h]
		\centering
		\begin{minipage}[c]{0.48\textwidth}
			\centering
			\includegraphics[width=7cm]{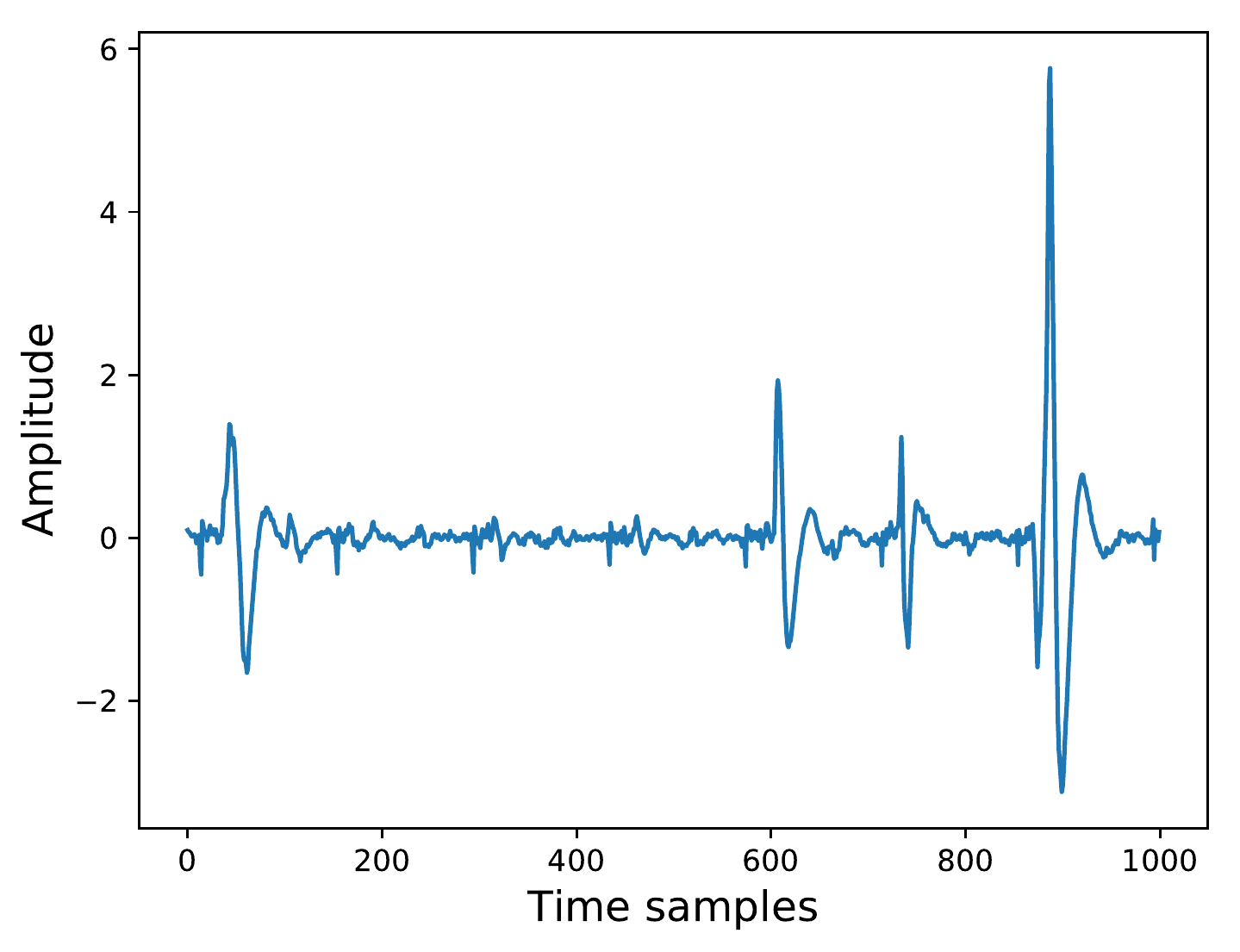}
		\end{minipage}
		\hspace{0.01\textwidth}
		\begin{minipage}[c]{0.48\textwidth}
			\centering
			\includegraphics[width=7cm]{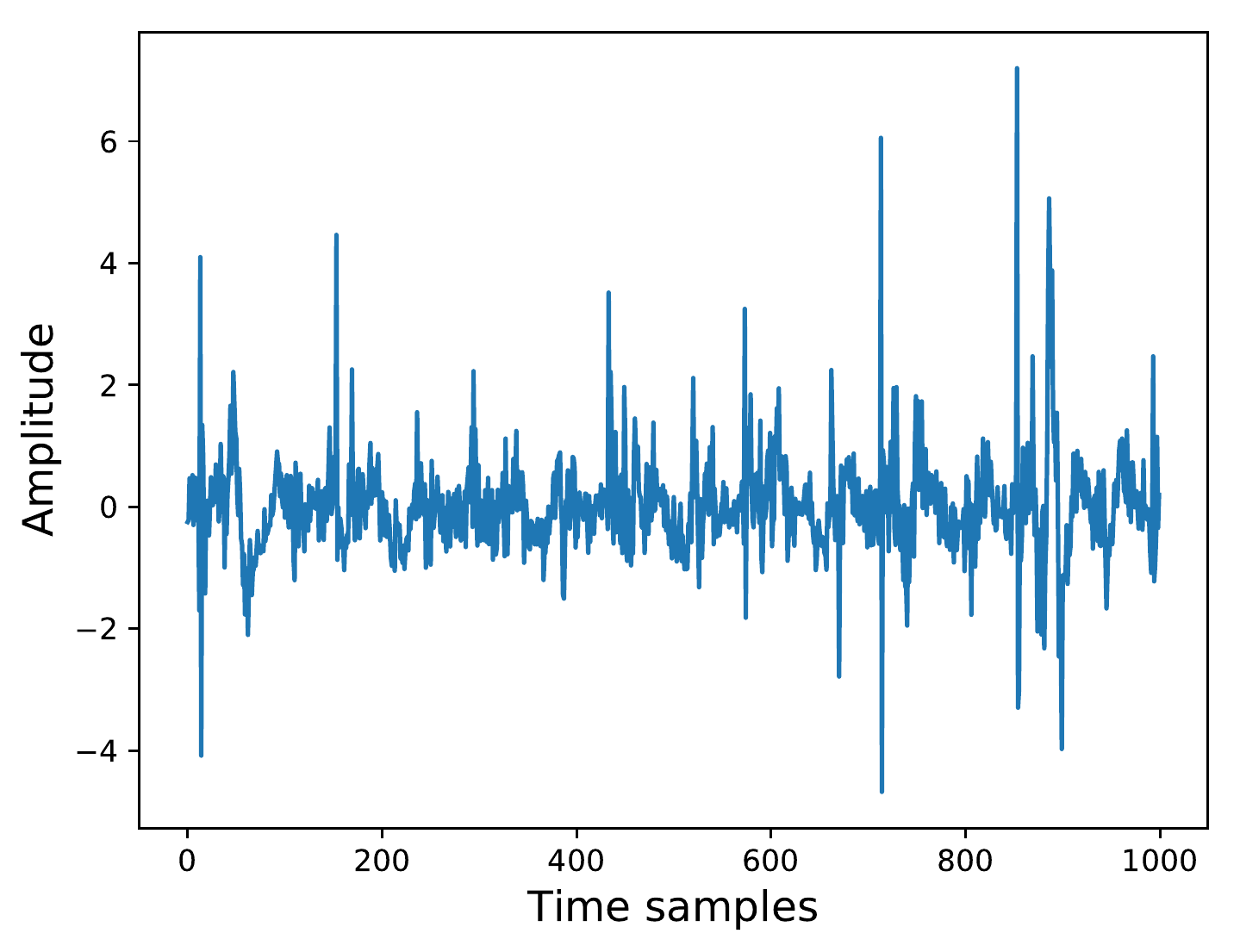}
		\end{minipage}
		\caption{Left: DPA of 1000 original traces. Right: DPA of 1000 generated traces.}
		\label{dpa_dpa}
	\end{figure}

\begin{figure}[h]
		\centering
		\begin{minipage}[c]{0.48\textwidth}
			\centering
			\includegraphics[width=7cm]{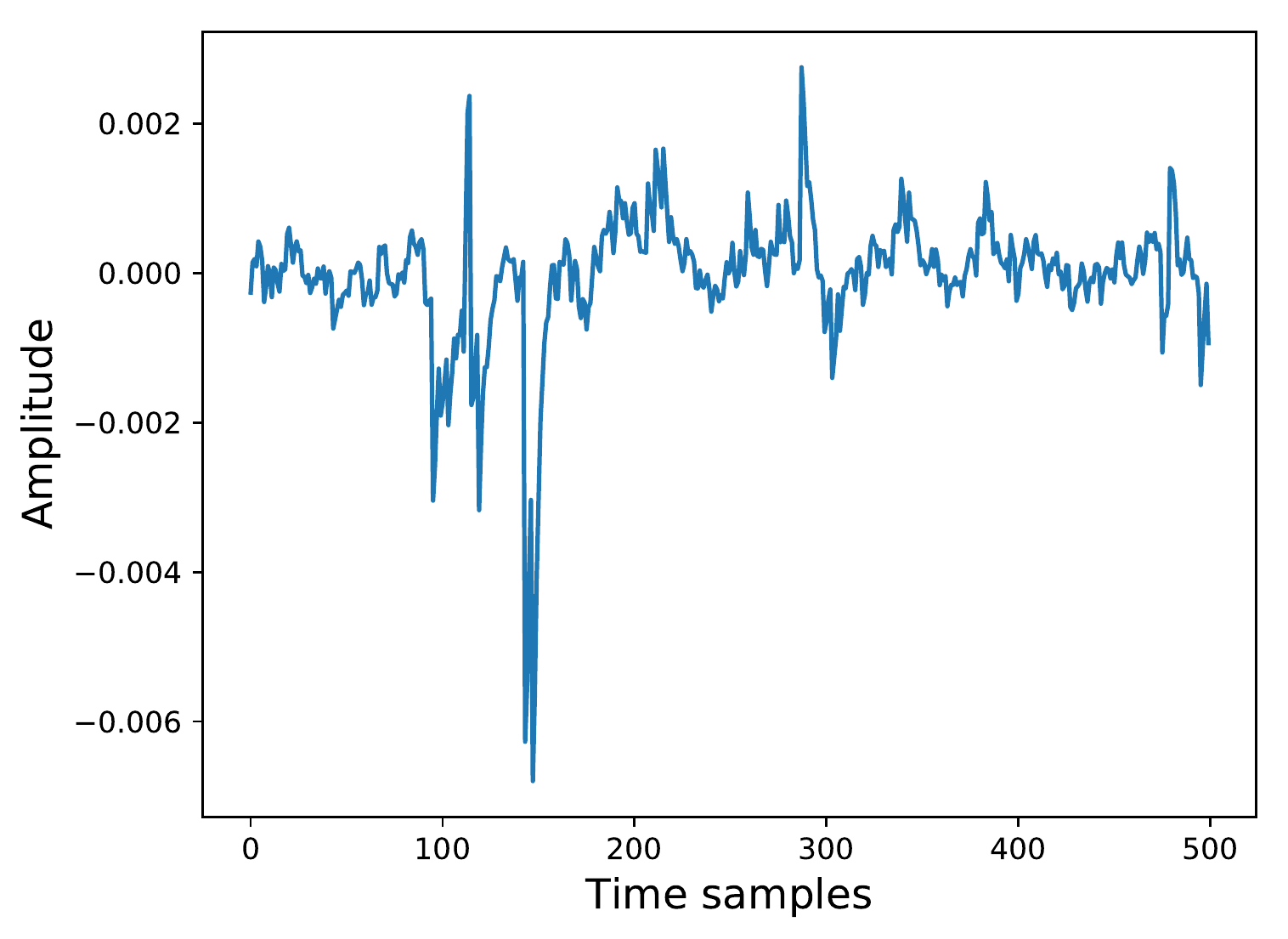}
		\end{minipage}
		\hspace{0.01\textwidth}
		\begin{minipage}[c]{0.48\textwidth}
			\centering
			\includegraphics[width=7cm]{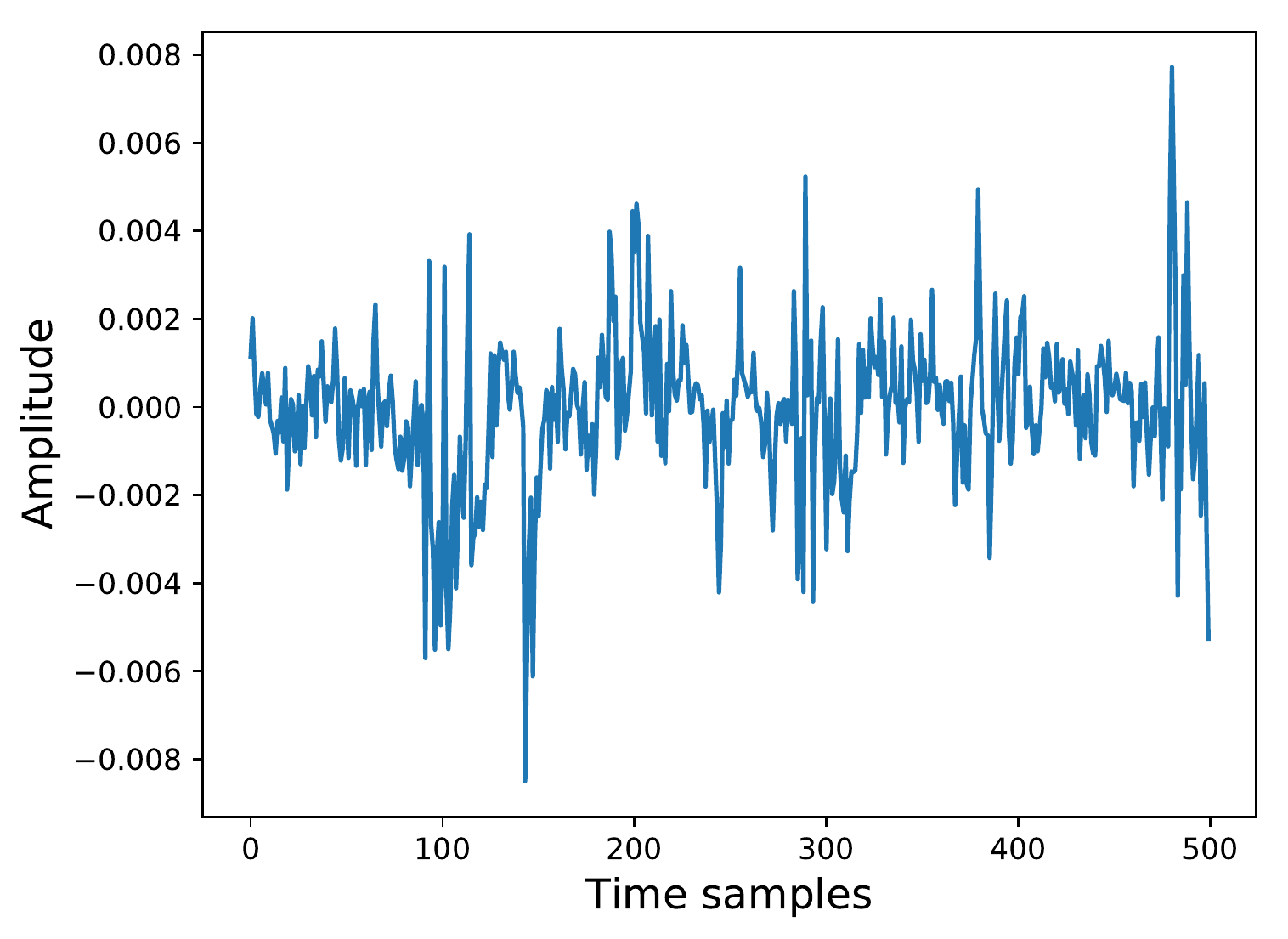}
		\end{minipage}
		\caption{Left: DPA of 100 original traces. Right: DPA of 400 generated traces.}
		\label{cw_dpa}
	\end{figure}

	From top to bottom, Figure \ref{sim_dpa}, Figure \ref{dpa_dpa} and Figure \ref{cw_dpa} correspond to the traces used for CPA analysis in \autoref{sec:sim}, \autoref{sec:dpav4} and \autoref{sec:CW} respectively.
	
	\subsection{The Shape of Generated Traces}
	\label{shape}
\begin{figure}[H]
		\centering
		\includegraphics[width=.5\textwidth]{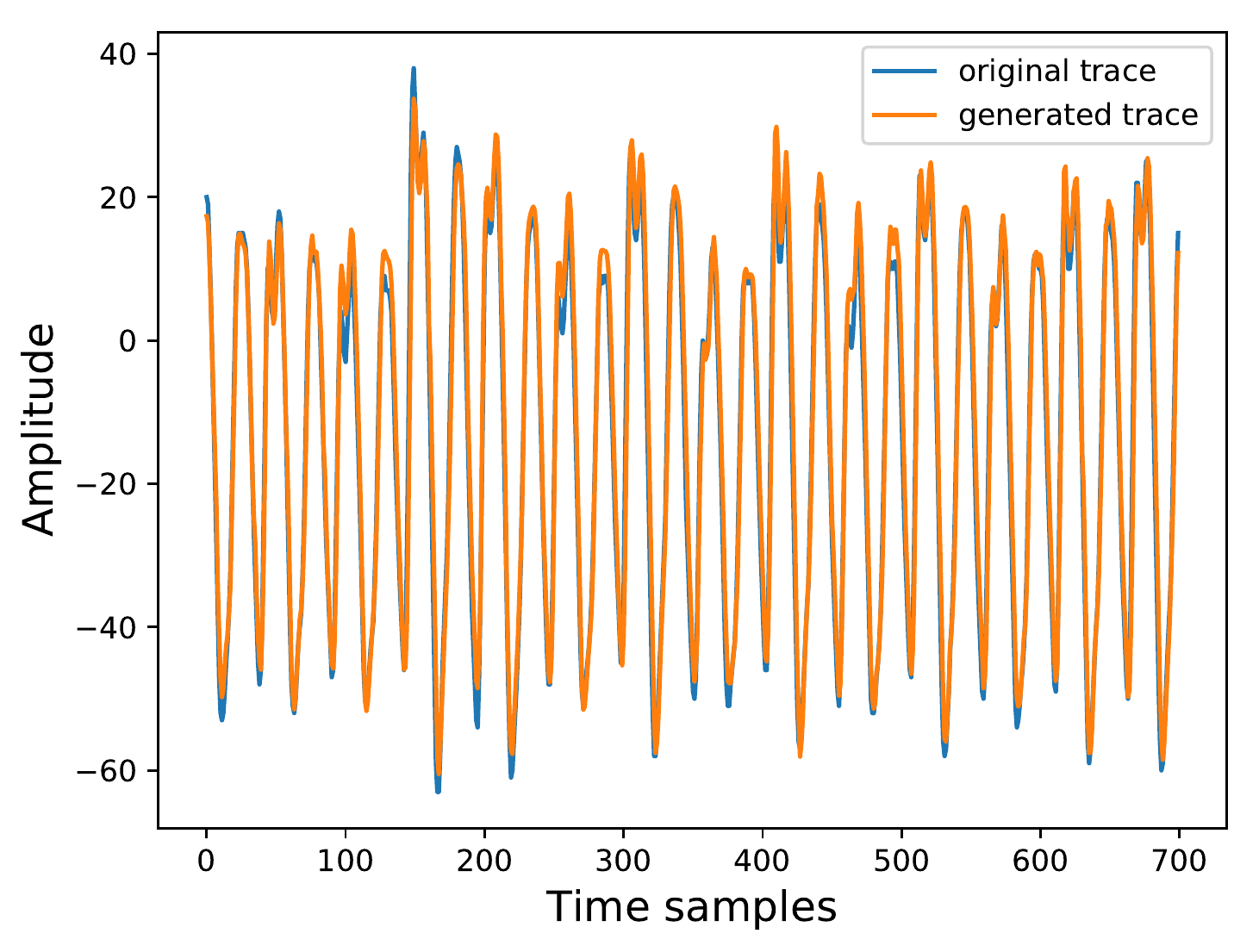}
		\caption{The shape of original trace and generated trace (ASCAD).}
		\label{ascad_shape}
	\end{figure}

	The shape of the generated trace is similar to the original one. The difference is that there are slight amplitude jitters at the peaks of the generated trace, and these slight jitters contain useful information we need for modeling. The comparison results are shown in Figure \ref{dpa_shape}, Figure \ref{cw_shape} and Figure \ref{ascad_shape}. 
	
	\noindent\begin{minipage}[t]{.48\textwidth}
		\centering
		\includegraphics[width=7cm]{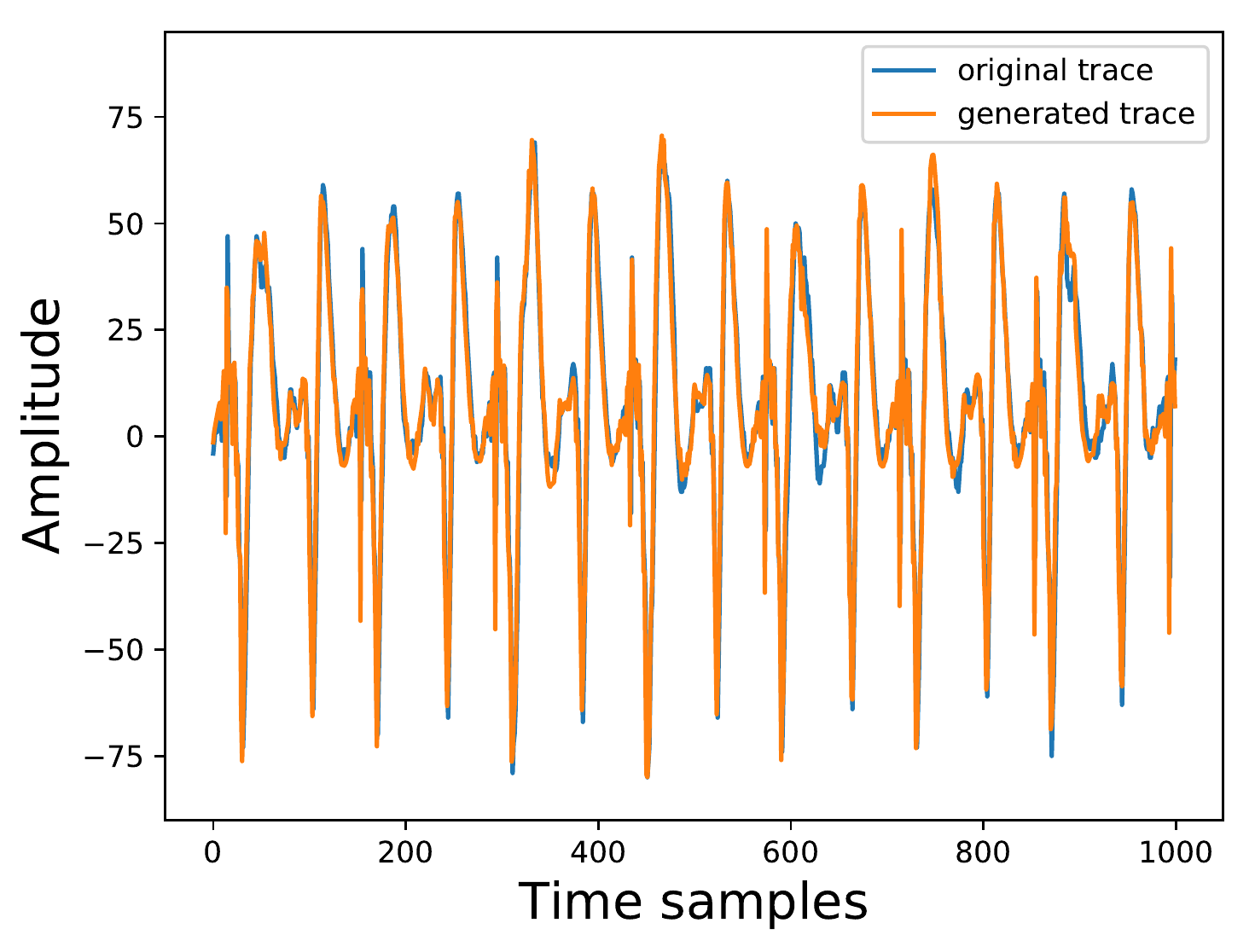}
		\captionof{figure}{The shape of original trace and generated trace (DPAv4).}
		\label{dpa_shape}            
	\end{minipage}%
	\hspace{0.02\textwidth}
	\begin{minipage}[t]{.48\textwidth}
		\centering
		\includegraphics[width=7cm]{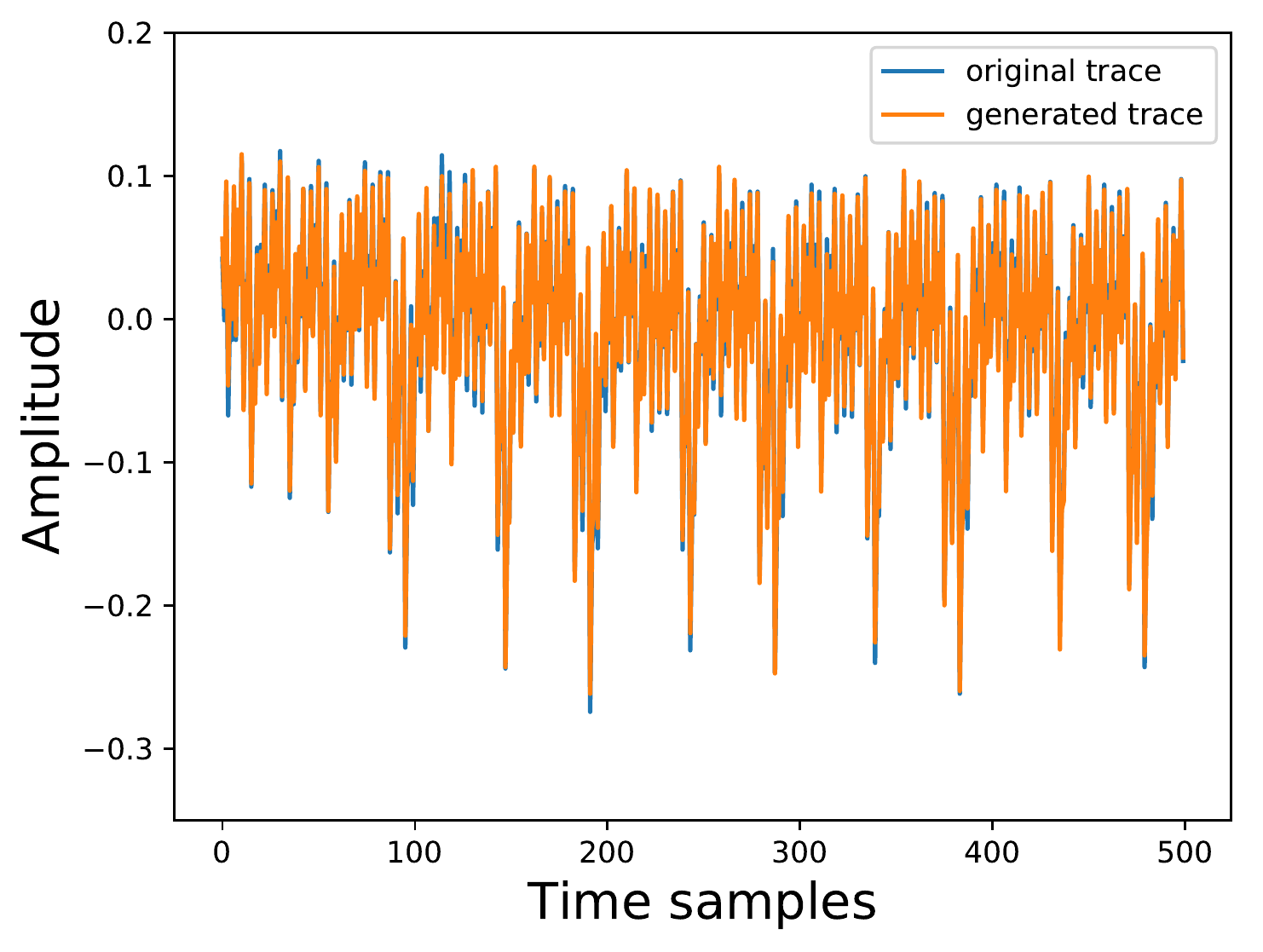}
		\captionof{figure}{The shape of original trace and generated trace (CW). \vspace{24pt}}
		\label{cw_shape}            
	\end{minipage}
	
	\subsection{Classifiers}
	\label{appendix_classifier}
	
	\begin{table}[h]
		\footnotesize
		\tabcolsep 10pt 
		\begin{tabular*}{\textwidth}{ccc}
			\toprule
			& \textbf{CW} & \textbf{ASCAD}\\
			\midrule
			\textbf{SVM} & \makecell[c]{Margin: $C=100$\\ kernel parameter: $\gamma=0.001$} & \makecell[c]{Margin: $C=500$\\ kernel parameter: $\gamma=0.001$}\\
			\midrule
			\textbf{RF} & \makecell[c]{The number of trees:\\$n\_estimators=100$} & \makecell[c]{The number of trees:\\$n\_estimators=1000$}\\
			\midrule
			\makecell[c]{\textbf{CNN}(filters, size, \\stride, padding, activation)} & \makecell[c]{(32, 11, 1, ‘same’, relu)\\(64, 11, 1, ‘same’, relu)\\Dense output with softmax} & \makecell{ (64, 11, 1, ‘same’, relu)			
				\\AveragePooling(size=2, stride=2)		
				\\(128, 11, 1, ‘same’, relu)		
				\\AveragePooling(size=2, stride=2)		
				\\(256, 11, 1, ‘same’, relu)		
				\\Dense(2048, relu)		
				\\Dense(2048, relu)			
				\\Dense output with softmax}\\
			\bottomrule
		\end{tabular*}
	\end{table}

\end{document}